**Two-Dimensional Magnetic Materials from Ferromagnetism to Altermagnetism: Fabrication, Symmetry, Evidence, and Device Readiness**

*Guojian Zhao, Ziang Meng*, Li Liu, Zhiyuan Duan, Jingyu Li, Sixu Jiang, Xiaoyang Tan, Shiheng Bu, Chen Liang, Xiaoning Wang, Peixin Qin*, Zhiqi Liu**

G. Zhao, Z. Meng, L. Liu, Z. Duan, J. Li, S. Jiang, X. Tan, S. Bu, C. Liang, P. Qin, Z. Liu
School of Materials Science and Engineering, Beihang University, Beijing 100191, China
E-mail: mengza@buaa.edu.cn; qinpeixin@buaa.edu.cn; zhiqi@buaa.edu.cn

G. Zhao, Z. Meng, L. Liu, Z. Duan, J. Li, S. Jiang, X. Tan, S. Bu, C. Liang, P. Qin, Z. Liu
State Key Laboratory of Tropic Ocean Engineering Materials and Materials Evaluation, Beihang University, Beijing 100191, China.

X. Wang
The Analysis & Testing Center, Beihang University, Beijing 100191, China.

**Abstract:** Atomically thin magnets are governed by composition, thickness, stacking, strain, defects, substrates, ferroic order, and interfaces. This Review organizes two-dimensional (2D) magnetism using net magnetization and momentum-space spin splitting or spin-polarization patterns as primary coordinates, supplemented by sublattice structure and the symmetry relations between opposite-spin sublattices. Ferromagnets generally carry a net moment; ferrimagnets can be fully compensated. Conventional antiferromagnets may retain spin-degenerate bands, while altermagnets permit nonrelativistic alternating spin splitting under dimension-specific symmetry constraints. M1–M4 reporting classes introduced here distinguish layer-resolved atomically thin forms, non-layer-resolved thin flakes or films, bulk benchmarks, and theoretical candidates or design routes; the evidence category and ordering-temperature provenance are reported independently. Current experimental support for altermagnetism comes mainly from M2 thin-film and M3 bulk benchmarks, leaving a fully verified atomically thin M1 altermagnetic phase as an open experimental goal. Most proposed monolayer and few-layer systems therefore fall within M4, and contested cases clarify the measurements needed for phase assignment. Device readiness is compared qualitatively using operating temperature, readout, switching, domain control, stability, scalability, interface sensitivity, and integration difficulty. Reliable phase assignment and device assessment require complementary measurements from the same material form under matched conditions, ideally on the same specimen or in situ.

## 1. Introduction

Atomically thin materials and van der Waals (vdW) assembly provide a platform for combining layered crystals into designed heterostructures [1,2]. In 2D magnetism, the central opportunity has expanded from demonstrating order at reduced dimensionality to engineering the magnetic symmetry of the active layer. The Mermin–Wagner theorem excludes spontaneous long-range magnetic order at nonzero temperature in ideal two-dimensional isotropic Heisenberg systems with short-range exchange [3-6]. Real 2D magnets can sustain ordered phases through magnetic anisotropy, discrete symmetry, interlayer coupling, substrate fields, and interfaces. These factors also provide readout channels and can alter both real-space net magnetization and spin-dependent bands in momentum space, thereby changing switching, domain behavior, and transport response [7-10].

The first direct demonstrations of intrinsic 2D magnetism in monolayer $CrI_3$ and few-layer $Cr_2Ge_2Te_6$ established the experimental basis for this field [11-12]. $Fe_3GeTe_2$ then connected atomically thin magnetism to itinerant metallic transport and gate-tunable ferromagnetism [13-14]. Subsequent work showed that electrostatic doping, vertical electric fields, pressure, strain, and stacking can switch or reshape layer-dependent magnetic states [15-21]. Tunneling measurements resolve field-dependent layer configurations in $CrI_3$ devices [22-24]. Assembly into vdW heterostructures and magnetic proximity can further reshape magnetic and valley states [25-26]. These advances motivate a comparison of the spintronic opportunities offered by different magnetic orders. Figure 1 outlines the Review architecture by placing ferromagnetic (FM), ferrimagnetic (FiM), antiferromagnetic (AFM), and altermagnetic (AM) orders at the center, and connecting them to synthesis and assembly, materials libraries, symmetry control, characterization and readout, and spintronic opportunities.

To compare these magnetic orders on common terms, this Review combines two primary organizing coordinates with a broader symmetry-based classification. The first, real-space net magnetization, distinguishes finite-moment and compensated states, with ferrimagnets spanning both. The second, momentum-space spin splitting or spin-polarization pattern, constrains whether a compensated state can act as a spin-polarized transport medium. A complete assignment combines these coordinates with the magnetic-sublattice arrangement and the symmetry operation relating opposite-spin sublattices, including whether that relation permits nonrelativistic momentum-dependent spin splitting. Together, the coordinates and symmetry test organize the four magnetic families emphasized here: ferromagnets, ferrimagnets, conventional antiferromagnets, and altermagnets [27-36]. Table 1 distinguishes altermagnetism from conventional antiferromagnetism and related spin-splitting mechanisms, Table 2 defines the minimum evidence package for 2D magnetic phase assignments, and Table 3 sets out the qualitative device-readiness considerations used later in the Review.

Ferromagnets combine a finite moment with exchange-split bands, enabling strong magneto-optical, anomalous Hall, tunneling, or spin-torque responses in suitable materials and device geometries. Ferrimagnets contain antiparallel, magnetically inequivalent sublattices; their net moment depends on sublattice balance and temperature and can vanish in a compensated state [30]. Many conventional collinear antiferromagnets have negligible net magnetization and retain spin-degenerate bands when combined symmetries protect the degeneracy. Altermagnets are collinear compensated magnets in which spin and crystal symmetries permit nonrelativistic, momentum-dependent alternating spin splitting [34-36].

The core question addressed here is therefore the following: when layer number, stacking, strain, defects, substrate coupling, ferroic order, and interfaces change magnetic symmetry, how do real-space compensation and momentum-space spin splitting together determine readout, transport, domain control, and device functionality? This question is especially important in 2D systems because the same perturbation that enables control can also change phase purity, stoichiometry, oxidation state, domain population, or surface symmetry [37-42].

Throughout this Review, representative entries in Tables 4, 5, 6A, and 6B receive an M1–M4 reporting class. M1 denotes an experimentally realized atomically thin material form with an explicitly reported layer number, including monolayer, bilayer, or discrete few-layer states under the stated field, gate, pressure, or device condition. M2 denotes a thickness-defined thin flake, film, or nanosheet for which a discrete atomic layer count has not been established; thickness, substrate, interface, and control conditions are reported explicitly. M3 denotes a layered-bulk, bulk, or thick-film sample used as a benchmark without a demonstrated thickness-dependent 2D state. M4 denotes an unrealized 2D material candidate, material-specific calculation, design route, or screening framework. Borderline M2/M3 entries report thickness and classification rationale. The M1–M4 labels record material form and realization status; they do not rank evidentiary strength. The evidence category is reported separately as a direct magnetic-order measurement, symmetry-sensitive or supporting measurement, indirect inference, or calculation; ordering-temperature provenance is reported as measured, inferred, or calculated.

The terms quasi-2D bulk and 2D magnetic lattice describe anisotropic dimensionality but do not by themselves establish an atomically thin phase.

Recent reviews have addressed 2D vdW magnetism [7-10,27-29,37,43], ferrimagnetic and antiferromagnetic spintronics [30-33], altermagnetism [34-36,44], and 2D magnetic and topological materials for low-power spintronics [45]. Building on these contributions, this Review asks a different question: how can magnetic states reported across atomically thin crystals, thickness-defined films, bulk benchmarks, and theoretical candidates be compared without conflating material form, evidence category, ordering-temperature provenance, and device readiness? The resulting framework integrates FM, FiM, AFM, and AM while linking each material form to symmetry-appropriate characterization and qualitative device assessment. Section 2 introduces fabrication and structural control shared across the magnetic families; Sections 3–5 examine their materials and magnetic states, followed by device comparisons and experimental priorities in Sections 6 and 7.

## 2. Fabrication, Structural Control, and Phase Verification of Two-Dimensional Magnetic Materials

Fabrication does more than control the shape and thickness of a 2D magnet; it can also determine the magnetic symmetry probed by a device. Layer number, stacking, strain, edge termination, defect density, oxidation state, stoichiometry, substrate charge transfer, and encapsulation can all modify exchange interactions and magnetic anisotropy. This section introduces these shared control variables; the magnetic structures and phase criteria of the representative materials are developed in Sections 3–5.

### 2.1. Candidate Identification, Exfoliation, Encapsulation, and van der Waals Assembly

Computational screening has become an important first step in the search for new 2D magnets. High-throughput exfoliation studies and materials databases identify compounds that may be isolated as monolayers, and model calculations estimate magnetic anisotropy, exchange interactions, and transition temperatures [46-50]. These methods reduce the number of candidates for experimental testing, while synthetic accessibility remains an independent constraint. A predicted monolayer may be inaccessible because of competing phases, volatile elements, oxidation, or a growth window that is too narrow to preserve the required stoichiometry.

Mechanical exfoliation remains a well-established route for studying intrinsic thickness-dependent magnetism because it minimizes structural damage and preserves the parent lattice. Bulk work on cleavable $CrCl_3$ and layered $VI_3$ provides structural and magnetic benchmarks for comparison with ultrathin forms [51-52]. Epitaxial growth has provided a monolayer $CrCl_3$ platform [53], while exfoliation has enabled monolayer or few-layer studies of $CrI_3$, $CrBr_3$, $Cr_2Ge_2Te_6$, and $Fe_3GeTe_2$, as well as ultrathin structural and transport studies of $CrSiTe_3$ [11-13,54-55]. Gate control in $Fe_3GeTe_2$ and tunneling through $CrI_3$ later broadened the accessible phase space [14,22]. The method remains low-throughput and offers limited control over flake position, geometry, and contact reproducibility.

For air-sensitive magnets, sample preparation directly affects the measured properties. Hexagonal boron nitride reduces substrate disorder and surface roughness [56], and one-dimensional edge-contact architectures provide low contact resistance and high electronic performance in encapsulated devices [57]. All-dry transfer limits solvent contamination, and hot pick-up assembly produces cleaner van der Waals interfaces [58-59]. These steps can determine whether a measured transition comes from the intended crystal or from a degraded, oxidized, or strained surface. Chromium halides are especially sensitive: $CrI_3$ undergoes rapid hydrolysis and oxidation under ambient conditions [38]. The vdW Lego-block concept, all-dry stamping, hot pick-up assembly, and encapsulated $CrI_3$ tunnel devices provide a practical sequence from assembly to protected magnetic heterostructures (Figure 2a–d).

Stacking can also change the magnetic ground state. In $CrI_3$, the ligand environment controls the magnetic anisotropy, while the sign and strength of the interlayer exchange depend strongly on the relative displacement of adjacent layers [16,60]. This sensitivity underlies the electrical and pressure control of few-layer states discussed in Section 3.1. Spin-filter tunneling, scanning magnetometry, domain imaging, magnon spectroscopy, and proximity-induced valley splitting have provided complementary views of these layer-dependent magnetic states [22,25,61-63].

### 2.2. Bottom-Up and Epitaxial Growth

Bottom-up growth is needed to move 2D magnetism beyond isolated exfoliated flakes. Molecular beam epitaxy (MBE), chemical vapor deposition (CVD), and related vapor-phase methods can control film thickness, crystal orientation, grain structure, and interface chemistry over much larger areas. Wafer-scale $Fe_3GeTe_2$, large-area $Fe_5GeTe_2$, and CVD-grown 1T-$CrTe_2$ illustrate the progress toward scalable metallic 2D magnets [64-66]. Layered antiferromagnetic semiconductors such as CrOCl and FeOCl have also been prepared as high-quality crystals [67-68]. In all of these systems, growth temperature, source ratio, substrate choice, and post-growth treatment determine the strain, texture, phase purity, vacancy concentration, and surface termination. These factors can influence the magnetic ground state as strongly as the film thickness. Representative MBE examples include $Fe_3GeTe_2$ and $Fe_5GeTe_2$ films (Figure 3a–d). For CVD-grown 1T-$CrTe_2$, optical microscopy, atomic force microscopy, Raman spectroscopy, and scanning transmission electron microscopy jointly resolve thickness, morphology, stability, and atomic structure (Figure 3e–i).

Monolayer $VSe_2$ was initially reported to be ferromagnetic at room temperature [39]; subsequent angle-resolved photoemission spectroscopy (ARPES) and calculations favored a nonmagnetic charge-density-wave state, questioning the intrinsic origin of the reported ferromagnetism [40]. This case highlights the need to relate magnetic signatures to the sample's phase and chemical state (Table 2).

Epitaxial growth also gives access to materials that cannot be obtained by cleaving a layered bulk crystal. Suitable substrates can stabilize metastable monolayers and crystal orientations that are unavailable through exfoliation. Confined and bottom-up methods have produced monolayer manganese selenides and one-unit-cell ferrite nanosheets [69-70]. Such approaches are particularly important for compensated and altermagnetic materials because many proposed oxides, chalcogenides, and fluorides are not naturally exfoliable. High-throughput searches for candidate 2D altermagnets therefore consider synthetic accessibility together with magnetic symmetry [71].

**2.3. Defects, Stoichiometry, Phase Control, and Characterization**

Defects are part of the magnetic problem in two dimensions. They change carrier density and valence, pin domains, modify anisotropy, and can introduce extrinsic ferromagnetic signals. Fe deficiency and site occupancy affect the ordering temperature, anisotropy, and moment of bulk $Fe_3GeTe_2$ [72-73]; self-intercalation in epitaxial $Fe_5GeTe_2$ offers a separate means of modifying exchange interactions [74]. Antisite disorder in $MnBi_{2n}Te_{3n+1}$ compounds simultaneously affects magnetism, carrier density, and topological transport [41]. In proposed altermagnets, twinning, surface reconstruction, and uncompensated surface moments can shape the measured response. The $RuO_2$ literature illustrates the dependence of classification on structure and sample quality [42]. Figure 4 illustrates $CrI_3$ degradation chemistry and compositional and structural features of Fe-Ge-Te compounds. These cases make chemical identity, surface condition, and domain population integral components of the magnetic evidence.

Structural, chemical, and magnetic characterization is best planned as a coordinated workflow. Atomic force microscopy and cross-sectional electron microscopy determine thickness and morphology; diffraction, Raman spectroscopy, and transmission electron microscopy establish phase, orientation, and stacking; and chemical analysis and core-level spectroscopy constrain stoichiometry and oxidation state. Magnetic order can then be tested with complementary probes suited to the material and thickness. Magneto-optics and Hall transport provide optical and electrical readout; superconducting quantum interference device (SQUID) magnetometry constrains net magnetization; Raman spin–phonon signatures provide indirect or supporting evidence; second-harmonic-generation domain imaging provides symmetry-sensitive evidence; and scanning single-spin and nitrogen-vacancy magnetometry provide local magnetic information [61,75-78]. Muon spin rotation, neutron scattering, and momentum-resolved spectroscopy are especially important when the order is compensated or when band spin splitting is central to the assignment [79-80].

### 2.4. Symmetry-Preserving Fabrication and Phase Verification of Compensated Magnets

For compensated magnets, growth conditions also determine whether the symmetry relation between opposite-spin sublattices is preserved. Moderate disorder may broaden the bands of a ferromagnet without eliminating its net moment. In an altermagnet, the relevant response depends on whether the full symmetry of the fabricated structure permits alternating spin splitting [81-83]. Section 5.1 explains the additional constraints in two dimensions. Domain configuration, surface termination, twin population, and Néel-vector orientation are essential characterization variables. They determine whether the predicted spin splitting, angle-dependent transport, or domain-dependent signal can be observed in the measured specimen.

The appropriate fabrication method depends on the property to be preserved or targeted. Mechanical exfoliation is well suited to intrinsic thickness-dependent studies. Epitaxy offers control over crystal orientation and interfaces. Van der Waals assembly is especially useful for electrostatic control, proximity exchange, spin-valley coupling, and interfacial spin–orbit torque. The external-control examples in Figure 5, including electrostatic gating, floating-gate-assisted nonvolatile control, pressure, and ionic gating, show that the target thickness alone does not determine the magnetic state.

## 3. Two-Dimensional Ferromagnets and Ferrimagnets

Ferromagnets and ferrimagnets away from compensation offer the most direct readout of 2D magnetic order. Their magnetic states can be probed by magneto-optical Kerr rotation, reflective magnetic circular dichroism, anomalous Hall transport, tunneling magnetoresistance, and magnetic imaging [27]. A finite moment simplifies readout but also introduces stray fields, device crosstalk, and domain formation at high device densities. Ferrimagnets can reduce the stray-field penalty through partial compensation while preserving electrical or optical readout over an appropriate temperature range. Table 4 summarizes the finite-moment, ferrimagnetic, and multisublattice systems discussed in this section.

### 3.1. Insulating and Semiconducting Ferromagnets

Chromium-based insulating and semiconducting magnets provided the first clear examples of intrinsic 2D magnetic order. $CrI_3$ is the standard monolayer Ising ferromagnet, and its layered parent crystal shows strong coupling between structure and magnetism [11,60,84]. $Cr_2Ge_2Te_6$, $CrBr_3$, $CrCl_3$, $CrSiTe_3$, and $VI_3$ span a broader range of exchange interactions, magnetic anisotropies, optical responses, and environmental stability [12,51-55]. Collectively, studies of these compounds established layer-dependent magnetic ground states, with monolayer optical detection demonstrated in selected members.

In chromium halides, the coupling between layers is as important as the order within each layer. A calculated stacking-dependent mechanism was reported for bilayer $CrI_3$ [16]. Electrostatic doping, vertical electric fields, and hydrostatic pressure can switch the balance between ferromagnetic and antiferromagnetic layer alignment in $CrI_3$ [17-21]. This sensitivity reflects changes in stacking, ligand geometry, interlayer spacing, and charge distribution. The same material can therefore act as a monolayer ferromagnet, an antiferromagnetically coupled few-layer spin filter, or an electrically tunable magnetic element. Material identity, stacking geometry, and the external control field jointly determine the accessible magnetic state.

Spin-filter tunneling is one of the most established device functions of insulating 2D magnets. In a vertical junction, the tunneling probability changes when successive magnetic layers switch between parallel and antiparallel spin-dependent barriers [23-24]. Scanning magnetometry has resolved layer-dependent magnetic fields in $CrI_3$ [61], direct imaging has revealed magnetic domains and moiré states in twisted structures [62], and spectroscopy has detected two-dimensional magnons [63]. In adjacent semiconductors, magnetic proximity provides a separate optical probe of interfacial exchange [26]. The discovery of intrinsic 2D-XY ferromagnetism also showed that atomically thin magnetic order is not limited to the Ising case [53]. The $CrI_3$ panels in Figure 6a–c, Figure 6g, and Figure 6h summarize optical imaging, Kerr-rotation mapping, and calculations of stacking-dependent interlayer exchange. Figure 6d–f presents complementary Kerr-rotation and domain-imaging results for $Cr_2Ge_2Te_6$.

The main practical constraints are low ordering temperature, environmental stability, and circuit integration. Most insulating 2D ferromagnets order far below room temperature [27], and some, notably $CrI_3$, degrade rapidly in air [38]. Their insulating character is useful for spin filtering and optical spectroscopy but makes them less suitable for planar current-driven circuits. Metallic ferromagnets provide a more direct path to electrical integration, while compensated magnets reduce the stray-field penalty in dense devices. Chromium halides therefore remain especially valuable as model systems for layer-dependent exchange, electrostatic control, and proximity effects.

### 3.2. Metallic and High-Temperature Ferromagnets

Metallic 2D ferromagnets connect atomically thin magnetic order directly to charge transport. $Fe_3GeTe_2$ is a widely studied itinerant van der Waals ferromagnet that combines metallic conduction, perpendicular magnetic anisotropy, and anomalous Hall readout [13,85], with ionic-gate control [14] and current-induced switching [86-87]. Because carrier density, exchange splitting, and anisotropy are closely linked, $Fe_3GeTe_2$ offers a more direct route to electrical integration in spin–orbit torque devices than insulating chromium halides.

$Fe_5GeTe_2$, $Fe_3GaTe_2$, and epitaxial $CrTe_2$ have shifted attention toward higher ordering temperatures [88-90]. Several compounds in these families approach or exceed room temperature in bulk, thin-film, or few-layer forms. Current-induced switching and orbital-torque switching in related van der Waals ferromagnets further show that high ordering temperatures can coexist with electrical control [91-92]. Reproducibility depends on Fe occupancy, phase composition, thickness, strain, and oxidation, all of which affect the transition temperature and anisotropy. Figure 7 collects representative $Fe_3GeTe_2$, $Fe_5GeTe_2$, $Fe_3GaTe_2$, and $CrTe_2$ data that illustrate the promise of high-temperature finite-moment order and the importance of specifying material form.

Metallic 2D ferromagnets offer strong electrical readout and current-driven control while retaining the magnetostatic drawbacks of ferromagnetism. Contact design governs spin injection [93]; Joule heating can complicate current-driven measurements [94]. Tunnel-barrier and van der Waals interface quality affect spin transport [95-96]. Stray-field crosstalk and domain-limited reproducibility are additional device-design considerations for dense integration. These compounds are among the more mature candidates for near-term devices, although high-density integration may favor lower-moment or compensated magnetic order.

### 3.3. Ferrimagnetism and Compensation

Ferrimagnets contain antiparallel, magnetically inequivalent sublattices. Their contributions generally leave a finite net moment, but can also cancel at a compensation point or within a fully compensated state. Magnetization compensation and angular-momentum compensation are distinct conditions, with different implications for readout, switching, and domain-wall dynamics [30]. This range of behavior connects finite-moment 2D platforms to compensated spintronics.

The set of experimentally reported 2D and ultrathin ferrimagnetic systems remains limited. One-unit-cell $Cr_2S_3$ and chemically synthesized $Cr_2S_3$ nanosheets were reported as ferrimagnetic on the basis of finite-moment measurements [97-98]. $CoFe_2O_4$ nanosheets have been synthesized down to one unit cell, while magnetic contrast and remanence were measured in thicker nanosheets [70]. The antiparallel sublattice configurations in these ultrathin forms have not yet been resolved directly. $Mn_3Si_2Te_6$ serves as a layered bulk ferrimagnetic benchmark with competing exchange and unusual magnetotransport [99-100]. Figure 8 illustrates these examples, and Table 4 links each material form to its reporting class and evidence category.

The Fe-Ge-Te and Fe-Ga-Te metallic families require material-specific assignments. Fe-site occupancy and deficiency affect $Fe_3GeTe_2$ [72-73], and self-intercalation modifies $Fe_5GeTe_2$ [74]. $Fe_3GaTe_2$ exhibits above-room-temperature ferromagnetism with large perpendicular anisotropy [89]. Multiple magnetic sites alone do not identify ferrimagnetism: the defining structural evidence is an antiparallel arrangement of magnetically inequivalent sublattices.

Fully compensated ferrimagnets extend this comparison to zero-moment states with spin-split bands. In the collinear, spin-conserving limit at zero temperature, the spin moment is proportional to the difference between the occupied spin-up and spin-down electron counts. At fixed integer filling, a gap in at least one spin channel can keep both counts integer; a state with equal counts can therefore retain exact compensation while its inequivalent sublattices produce different spin-resolved bands [101]. This filling-based mechanism differs from symmetry-enforced compensation in an altermagnet. The associated *s*-wave description denotes spin splitting without the symmetry-required sign alternation characteristic of altermagnetism; it does not imply a momentum-independent splitting.

Proposed 2D routes include staggered potentials, Janus structures, and alloying [101], an antiferromagnetically coupled $CrI_3/CrGeTe_3$ heterostructure [102], and spin-order control in B-stacked $Cr_2C_2S_6$ bilayers [103]. In the latter model, reversing one layer's Néel vector produces a compensated ferrimagnetic configuration above a ground state that preserves combined inversion and time-reversal (*PT*) symmetry. These calculations illustrate how the spin channels can remain distinct despite a vanishing net spin moment, with implications for valley-selective electronic states [104]. An experimental test would need to establish the antiparallel sublattice order, compensation, and band structure in the same material form.

When both spin channels cross the Fermi level, the gap-based electron-counting argument no longer applies. Calculations predict a fully compensated ferrimagnetic metal in monolayer $Cr_2O$ when an out-of-plane electric field or in-plane uniaxial strain breaks the fourfold rotoinversion $S_{4z}$ that relates the opposite-spin Cr sites [105]. A second route closes the gap of a bilayer built from unipolar magnetic semiconductors with antiferromagnetic interlayer coupling. Calculations illustrate this electric-field-driven route in MnOF, where reduced interlayer spacing stabilizes the required antiferromagnetic coupling [106]. These metallic states depend on the electronic filling and control conditions of each model, with finite-temperature stability and spin–orbit coupling (SOC) effects requiring further material-specific assessment.

## 4. Two-Dimensional Antiferromagnets

A 2D antiferromagnet differs from a ferromagnet in more than its net moment. In ideally compensated antiferromagnetic forms, the near-zero net moment often makes order harder to detect directly and generally suppresses stray fields. Strong exchange interactions set a fast dynamical scale; symmetry-sensitive optical signals and magnon-mediated transport offer complementary ways to study antiferromagnetic order [31-33]. Table 5 separates the representative antiferromagnetic 2D and quasi-2D materials by dimensionality, ordering-temperature provenance, and reporting class, with material-specific notes.

Three material classes organize this section. Transition-metal phosphorus trichalcogenides ($MPX_3$) show how magnetic order couples to phonons, magnons, and excitons [107-108]. CrSBr illustrates layer-dependent low-symmetry magnetic semiconductors: second-harmonic generation (SHG) supports a ferromagnetic monolayer state, while multilayers and bulk can realize A-type antiferromagnetic configurations with optical, transport, and magnetoelastic readout channels [109-111]. $MnBi_2Te_4$ connects bulk A-type magnetic order [112] with layer-parity- and field-dependent transport in few-layer films [113].

### 4.1. $MPX_3$ Antiferromagnets: Magnetic Order, Magnons, and Excitons

Transition-metal phosphorus trichalcogenides ($MPX_3$) constitute a well-established family of van der Waals antiferromagnetic semiconductors. $MnPS_3$, $FePS_3$, and $NiPS_3$ occupy different exchange and anisotropy regimes. $MnPS_3$ is close to a Heisenberg antiferromagnet, $FePS_3$ is a quasi-2D Ising system with zigzag order, and $NiPS_3$ has easy-plane or XXZ-like behavior. Neutron studies provide bulk reference points: spin waves and exchange parameters in $MnPS_3$ [114], the magnetic structure of $NiPS_3$ [115], and the spin-wave Hamiltonian of $FePS_3$ [116], whose analysis revised an earlier propagation-vector assignment [117]. These results guide the interpretation of few-layer and monolayer measurements.

The thickness dependence of $MPX_3$ compounds highlights the gap between exfoliability and robust monolayer order. Raman zone-folding signatures support Ising-like antiferromagnetic order down to monolayer $FePS_3$ [107], whereas Raman signatures of order persist to bilayer $NiPS_3$ but are strongly suppressed in the monolayer [118]. Complementary Raman studies probe thin-layer $MnPS_3$ [75] and identify a magnon in thicker $FePS_3$ flakes [76]. Second-harmonic imaging resolves antiferromagnetic domains and probes magnetic symmetry in bilayer and thicker $MnPS_3$, while the monolayer response is strongly suppressed [77]; optical linear dichroism provides symmetry-sensitive evidence in $FePS_3$ [119]. Figure 9 connects the structural motifs of $MPX_3$ compounds with representative optical probes of compensated magnetic order in $NiPS_3$ and $FePS_3$.

$MPX_3$ compounds also host correlated optical and magnetic excitations. In $NiPS_3$, a coherent many-body exciton appears below the Néel temperature, and resonant inelastic X-ray scattering reveals Hund's excitons whose dispersion follows the magnetic correlations [108,120]. Pressure changes the exchange pathways in $FePS_3$ and drives coupled magnetic and electronic phases [121]. The $MPX_3$ family is therefore a model system for correlated antiferromagnetic semiconductors, although its ordering temperatures remain modest.

### 4.2. Low-Symmetry Magnetic Semiconductors: Layer-Dependent Order and Anisotropy

In low-symmetry magnetic semiconductors, magnetic anisotropy becomes a useful control parameter. SHG measurements support a ferromagnetic state in isolated monolayer CrSBr and yield an inferred $T_C = 146 \pm 2$ K [111], whereas multilayers and bulk contain ferromagnetic CrSBr layers coupled antiferromagnetically along the stacking direction [109,122-123]. Even-layer states are compensated, whereas odd-layer multilayers retain an uncompensated layer moment while preserving A-type interlayer order. The layered magnetic order produces large negative magnetoresistance, and field-driven spin reorientation changes the electronic coupling between layers [109,124]. The crystal structure, cleaving behavior, and bulk temperature-dependent conductance and susceptibility provide baseline evidence for this air-stable anisotropic magnetic semiconductor (Figure 10a–c).

CrOCl illustrates the coupling between magnetic order and lattice distortion in low-symmetry oxyhalides. Early work established its bulk synthesis and neutron-determined magnetic structure [125]. Subsequent studies examined bulk magnetoelastic phases [126], calculated ground states in bulk and monolayer chromium oxyhalides [127], and spin-lattice coupling in CrOCl [128]. Layer-resolved tunneling reveals metamagnetic transitions [129]. CrOCl interfaces support correlated insulating states in adjacent bilayer graphene [130]. Multistate magnetoelectric storage in CrOCl further connects magnetic and structural control to nonvolatile operation [131].

Bulk and few-layer CrSBr, including compensated even-layer and uncompensated odd-layer stacks, provide a clear example of opto-spintronic functionality in a 2D magnetic semiconductor. Studies of dynamic magnetic crossover, layer-dependent spin reorientation, intrinsic magnetic behavior, and spin-filter tunneling provide complementary optical, magnetotransport, and vertical-tunneling probes of the interlayer antiferromagnetic state [111,122-123,132-134]. Tunable exciton–magnon coupling, self-hybridized polaritonic magneto-optics, ultrastrong exciton–polariton coupling, and magnon-mediated exciton–exciton interactions convert magnetic order into excitonic, polaritonic, and nonlinear optical signals [110,135-137]. This optical sensitivity is especially useful when the anomalous Hall response is weak or absent. Coherent and hybridized magnon modes, together with calculated exciton-energy modulation, illustrate the exciton–magnon coupling channel (Figure 10d–f). Layer-dependent magnetoresistance and magnetophotoluminescence demonstrate complementary electrical and optical readout (Figure 10g, h).

Recent studies have extended CrSBr from passive readout to nonvolatile control. A CrSBr-based floating-gate architecture incorporating hexagonal boron nitride, graphene, and $CuInP_2S_6$ uses persistent charge accumulation or depletion to produce nonvolatile changes in resistance, magnetoresistance, and the metamagnetic switching field [138]. The responses induced by tuning the electronic state, interlayer exchange, or excitonic environment provide indirect but sensitive probes of the magnetic order. Nitrogen-vacancy magnetometry of CrSBr [78] and electrically enhanced second-harmonic domain contrast in $MnPS_3$ [139] give complementary information about local magnetic fields and magnetic symmetry.

### 4.3. Layered Antiferromagnetic Topological Materials

$MnBi_2Te_4$ connects A-type magnetic order with nontrivial band topology. Each $MnBi_2Te_4$ septuple layer is ferromagnetic in plane, while neighboring septuple layers couple antiferromagnetically. Bulk measurements establish the magnetic and electronic reference state of $MnBi_2Te_4$ [112]. Thin-film predictions [140], epitaxial-film measurements [141], and layer-resolved calculations [142] show how its properties change with thickness. Layer parity, surface magnetization, magnetic field, and electrostatic tuning therefore help determine whether a film hosts an axion-insulator-like, Chern-insulator, quantum anomalous Hall, or conventional magnetic semiconductor state. The septuple-layer structure, exchange interactions, band structures, and topological phase diagrams of $MnBi_2Te_4$ provide the microscopic basis for this coupling between magnetism and topology (Figure 11a, b).

In $MnBi_2Te_4$ films, layer parity and magnetic field shape predicted and observed quantum anomalous Hall (QAH), axion-insulator-like, and high-Chern-number regimes [113,143-146]. Odd-layer films can host finite uncompensated magnetization and QAH behavior, even-layer films form natural platforms for compensated axion-like or zero-Hall regimes, and field-polarized states can enter Chern regimes. Reproducibility therefore depends on coordinated control of antisite disorder, magnetic domains, finite-size hybridization, surface degradation, Fermi-level drift, and contact processing, each of which can determine whether quantized transport is observed. Gate-dependent and field-dependent transport in five- and six-septuple-layer $MnBi_2Te_4$ devices illustrates how the quantum anomalous Hall, axion-insulator, and Chern-insulator regimes are selected experimentally (Figure 11c–f).

These materials illustrate several routes to detecting compensated order. The following section examines the additional symmetry conditions under which such order supports alternating spin splitting.

## 5. Toward Two-Dimensional Altermagnetism: Symmetry, Evidence, and Candidate Platforms

### 5.1. Symmetry Criteria and Practical Identification of Altermagnetism

Altermagnetism describes compensated collinear order whose nonrelativistic symmetry permits momentum-dependent alternating spin splitting [34-36,147-152]. A spin space group combines operations in spin and real space, allowing their roles to be examined separately. A 180° spin-space rotation about an axis perpendicular to the collinear spin direction reverses the spins and is unitary, unlike antiunitary time reversal; a sublattice-exchanging symmetry pairs it with an appropriate real-space operation. The relevant question is whether the complete group protects degeneracy between opposite-spin bands at every momentum or permits an alternating spin-polarization pattern away from protected nodes [81-83].

Two dimensions impose additional symmetry constraints. For a collinear layer in the *xy* plane without SOC, a sublattice-exchanging operation pairing a 180° spin rotation with the mirror $M_z$ enforces opposite-spin degeneracy at each in-plane momentum. The analogous operation involving $C_{2z}$ provides the same protection when combined with time reversal and a 180° spin rotation about an axis perpendicular to the spin direction. This auxiliary antiunitary symmetry preserves the spin channel and reverses momentum. In three dimensions, these relations can reverse $k_z$, so their degeneracy protection alone is restricted to invariant momentum planes. The full spin layer group therefore determines whether alternating spin splitting is allowed. In the type IV category of this classification, calculated monolayer $Hf_2S$ retains nonrelativistic spin-degenerate bands but permits an anomalous Hall effect (AHE) with SOC [153].

Spin space groups also distinguish magnetic structures whose net spin magnetization is symmetry-enforced to vanish from structures in which a finite value is allowed. Ferrimagnets belong to the latter category, including states with fully compensated spin magnetization. Oriented spin space groups fix the magnetic orientation and connect this exchange-symmetry description to the magnetic space group retained with SOC [154]. With SOC, the actual orientation and antiunitary symmetries must therefore be examined separately. For spinful electrons with SOC, *PT* symmetry protects double degeneracy throughout the Brillouin zone, whereas time reversal combined with a nonprimitive translation alone does not generally do so [34-36].

Altermagnetic order can coexist with a small symmetry-allowed ferromagnetic moment once SOC is included, as described by Landau analysis of MnTe [155]. In bulk MnTe, magnetometry and transport measurements revealed a weak ferromagnetic-like signal accompanying the anomalous Hall effect and vanishing at the Néel temperature [156]. For monolayer $RuF_4$, calculations predict SOC-induced canting and a weak net moment [157]. The dominant order, the weak moment, and their contributions to the measured signal should be identified separately. Rashba and Dresselhaus splittings originate from SOC, whereas altermagnetic splitting is permitted in the nonrelativistic reference state. Noncollinear antiferromagnets and collinear altermagnets can both exhibit Berry-curvature-driven transport, and magnetic multipoles provide useful descriptions of responses in both settings [149,155,158]. Their distinction rests on magnetic structure and symmetry; neither an anomalous Hall response nor a multipolar description uniquely identifies either class.

Practical identification therefore starts with the magnetic structure in the measured material form, followed by its full symmetry and the domain or Néel-vector orientation. Magnetic-structure databases provide a useful starting point [159]. Spin-integrated ARPES maps band dispersions and their momentum-dependent splitting, while spin-resolved ARPES additionally probes the spin polarization of the resolved bands. Transport, thermoelectric, and terahertz measurements supply complementary symmetry-sensitive information. Table 1 summarizes the distinctions, and Section 5.5 gives the corresponding experimental workflow.

### 5.2. Experimental Benchmarks, Candidates, and Contested Diagnostic Cases

Results on $RuO_2$ differ across probes and sample forms. Reports include tilted spin currents, spin-splitter torques, anomalous Hall response, time-reversal-symmetry-breaking ARPES signatures, and transport in single-variant $RuO_2$(101) thin films [79,160-164]. These studies establish strong sample-specific responses, while the corresponding bulk magnetic structure remains unresolved.

Muon-spin and neutron measurements did not detect long-range magnetic order [80], and ARPES measurements did not detect altermagnetic spin splitting [165]. A magnetic-structure reassessment [166] and time-domain terahertz measurements on $RuO_2$/permalloy bilayers, which found no inverse altermagnetic spin-splitting effect under the tested orientations and conditions [167], further constrain the interpretation. $RuO_2$ therefore remains a contested M2/M3 diagnostic case whose interpretation depends on material form, strain, surface condition, orientation, and probe.

CrSb and MnTe provide experimentally advanced M2/M3 benchmarks, with complementary measurements performed on different material forms. In CrSb, spin-integrated soft-X-ray ARPES (SX-ARPES) on 30 nm epitaxial films resolves a symmetry-consistent splitting of approximately 0.6 eV but does not determine spin polarization [168]. Bulk three-dimensional ARPES and spin-resolved ARPES report maximum splittings of approximately 0.93–1.0 eV and a reproducible spin-polarization contrast consistent with the predicted *g*-wave texture; the measured polarization may be reduced by domain averaging and momentum broadening [169-170]. MnTe evidence combines domain imaging and control in a 30 nm film [171], SX-ARPES on a 200 nm film [172], and bulk-crystal spin-integrated and spin-resolved ARPES [172-173].

Layered M3 compounds reveal how interlayer order can change an altermagnetic assignment. $KV_2Se_2O$ was reported as a room-temperature *d*-wave altermagnet using magnetic characterization and spin-resolved photoemission [174]. Subsequent neutron diffraction found G-type order in both bulk single crystals and polycrystalline powder, with $T_N \approx 400$ K determined from the powder data [175]. This structure is incompatible with the proposed bulk altermagnetic state. These results leave $KV_2Se_2O$ a contested diagnostic case.

$Rb_{1-\delta}V_2Te_2O$ presents a related, sample-dependent comparison. Single-crystal measurements reported a transition near 307 K, crystal-symmetry-paired spin–valley locking in spin-resolved photoemission, and suppressed intervalley scattering in quasiparticle interference [176]. A later powder neutron study found G-type antiferromagnetic order with $T_N = 337$ K [177]. The two temperatures and evidence sets describe different samples. Relating their magnetic structures to the electronic measurements requires control of composition and interlayer order.

Hidden altermagnetism offers a layer-resolved interpretation of compensated bulk structures. In a *PT*-symmetric stack, opposite layers can carry alternating spin textures that cancel in the globally spin-degenerate bands [178]. A recent preprint reports G-type order below approximately 257 K in bulk $Cs_{1-\delta}V_2Te_2O$ crystals with $\delta \approx 0.23$. Neutron diffraction establishes the G-type magnetic structure. Surface-sensitive spin-resolved photoemission preferentially samples near-surface sectors and detects local spin polarization consistent with layer-projected calculations [179]. For stoichiometric $CsV_2Te_2O$, calculations predict that uniaxial strain induces a net moment in the C-type configuration but not in the G-type configuration, offering a proposed way to distinguish them [180]. These results connect bulk stacking to layer-local electronic states, which remain distinct from globally spin-split altermagnetism.

Chromium analogues provide a complementary route. Calculations predict robust *d*-wave altermagnetic states in the $XCr_2Y_2O$ family ($X$ = K, Rb, Cs; $Y$ = S, Se, Te) [181]. Bulk $CsCr_2S_2O$ has now been synthesized, with powder neutron diffraction establishing C-type antiferromagnetic order below 326 K and transport showing a metal-to-insulator transition near 305 K. Its electronic spin splitting is calculated [182]. This is an M3 experimental magnetic-structure benchmark, with atomically thin realization and direct electronic spin-polarization measurements as the next tests.

$La_2O_3Mn_2Se_2$ is a correlated layered bulk system with measured antiferromagnetic order, two-dimensional short-range correlations, and calculated *d*-wave spin–momentum locking [183]. Modeling reproduces its powder inelastic-neutron-scattering spectrum using inequivalent second-neighbor exchange interactions characteristic of the *d*-wave spin Hamiltonian [184]. Here, two-dimensional refers to the magnetic lattice and exchange physics; the evidence establishes a layered M3 benchmark, while an atomically thin form and direct electronic spin-polarization measurements remain outstanding.

One-unit-cell epitaxial GdAlSi provides an experimentally realized M1 material form. Thickness-dependent transport revealed a spontaneous anomalous Hall response without a discernible net moment; calculations retained nonrelativistic spin splitting (NRSS), although dimensional reduction removed the model's defining fourfold spin-space operation and placed the one-unit-cell structure outside the altermagnetic class [185]. Bulk polarization-resolved resonant elastic X-ray scattering subsequently established an incommensurate cycloidal ground state [186]. GdAlSi therefore provides an M1 boundary case for separating NRSS from altermagnetic order. Direct measurements of the film's magnetic structure and spin polarization would further test its magnetic assignment.

Section 5.6 discusses additional dichroism, magnon, transport, and domain-control evidence for MnTe, CrSb, $RuO_2$, and related benchmark systems. Table 6 and Figure 12 list their M2/M3 material forms alongside the corresponding evidence categories.

Recent classification and design studies of two-dimensional altermagnets [187] inform Table 7, which groups material-specific candidates, symmetry-engineering routes, and screening frameworks as M4 entries. For material-specific candidates, experimental synthesis and thickness determination would establish an M1 or M2 material form; magnetic-structure and band-sensitive measurements would test the proposed altermagnetic phase.

**5.3. Electronic, Topological, and Correlation Effects**

Beyond symmetry classification, theoretical work addresses order parameters, topology, and electronic correlations. Landau theories identify altermagnetic order parameters and their symmetry-allowed couplings to secondary magnetic multipoles, weak magnetization, and observable responses [155]. Orbital-order models further show that altermagnetism can emerge from coexisting staggered orbital and antiferromagnetic order [188]. Predictions of topological transitions [189], neutron-scattering signatures [190], and correlation-driven anomalous Hall effects [191] connect altermagnetic order to both quasiparticle bands and collective excitations. Extending these models to realistic 2D structures calls for tests of strain, substrate fields, screening, and electrostatic doping.

Proposed altermagnetic systems extend these ideas to superconductivity and topological bands. Josephson-junction proposals couple an altermagnetic region to conventional superconducting electrodes [192]. Intrinsic topological superconductivity is also predicted for two-dimensional altermagnetic metals with attractive interactions and Rashba spin–orbit coupling [193]. Models with mirror Chern bands in two dimensions and Weyl nodal loops in three dimensions provide intrinsic topological routes set by altermagnetic crystal symmetry [194].

Valley control illustrates two distinct mechanisms. In tetragonal altermagnets with SOC included, reorienting the Néel vector can break the symmetry relating the valleys and split their energies [195]. In spin–layer-coupled systems, a perpendicular electric field acts on the layer degree of freedom and can tune valley energies through valley–layer locking; calculations for twisted VOBr bilayers and monolayer $Ca(CoN)_2$ provide examples [196]. The first mechanism depends on magnetic orientation and SOC, while the second exploits layer polarization. Where gating is used, changes in magnetic symmetry should be assessed alongside the valley response.

### 5.4. Candidate Platforms and Design Strategies for Two-Dimensional Altermagnetism

The M4 design space (Figure 13) includes material-specific monolayer predictions, stacking and twisting routes, polarity or ferroic control, and high-throughput or artificial intelligence (AI)-assisted screening.

Theoretical monolayer platforms include CrO [197] and vanadium tetrahalides [198], whose calculated magnetic structures permit nonrelativistic alternating spin splitting. $RuF_4$ also illustrates predicted coexistence with a weak ferromagnetic moment [157].

Chromium trihalides offer a theoretical route to altermagnetism starting from synthesized ferromagnetic monolayers. Calculations for $CrX_3$ ($X$ = Cl, Br, I) predict that biaxial compression stabilizes antiferromagnetic order, after which a perpendicular electric field or Janus substitution can break inversion while retaining a vertical mirror and permit *i*-wave altermagnetism [199]. In the reported calculations using the generalized gradient approximation, $CrCl_3$ requires the smallest compressive strain among the three parents, although the threshold depends on the treatment of electronic correlations.

Stacking, twisting, and spin-layer coupling offer complementary control routes. Bilayer models show how registry and twist alter the symmetry relating opposite-spin sublattices [200-202]. In calculated monolayer $Ca(CoN)_2$, spin-layer coupling instead enables gate-field control of spin polarization [203]. These approaches exploit the connection between crystal symmetry and momentum-space spin polarization [204-205].

Polarity and ferroic order broaden the M4 design space. Calculated routes include polar and Janus monolayers [206-208], electric-field-induced switching [209], and ferroelectric control of electronic and magnonic states [210]. Theoretical designs based on bulk ferroelectric magnets provide a general symmetry-based precedent [211]. Further proposals exploit antiferroelectric order [212], ferroelectric distortions [213], ferroelastic reorientation [214], spin-ferroelectric locking [215], and interlayer sliding [216]. A distinct multiferroic model supports hidden, sector-resolved altermagnetic polarization while retaining globally spin-degenerate bands [217]. AI-assisted altermagnet discovery [218], general 2D magnetic-order screening [219], and data-driven anisotropy studies [220] complement these symmetry-guided searches [71,187].

Among M4 candidates, experimental priority depends on more than the magnitude of the calculated spin splitting. High-priority systems combine a spin-split band near the Fermi level or an accessible band edge with magnetic anisotropy sufficient for two-dimensional order. They also require plausible exfoliation or epitaxial growth, chemical and air stability, a substrate or encapsulation strategy that preserves the target symmetry, and a realistic route to band-sensitive verification. Predictions whose transition temperature or spin splitting depend strongly on a chosen Hubbard $U$, imposed strain, ideal termination, or untested ferroic configuration should state that dependence explicitly. Measurements that resolve weak-moment contributions, SOC-driven splitting, and interface-induced responses can then identify the origin of the observed signal.

### 5.5. Evidence Standards for Two-Dimensional Altermagnetic Assignments

The reporting framework in Table 2 distinguishes theoretical candidates from experimentally realized material forms. For a material-specific M4 candidate, the phase assignment rests on a compensated collinear nonrelativistic reference state and a complete spin layer group analysis that permits nonrelativistic alternating spin splitting. An orientation-specific magnetic symmetry analysis is additionally needed when SOC is included. Assessing a candidate's feasibility also involves structural stability and magnetic anisotropy or another physically plausible mechanism for finite-temperature 2D order [71,81-83,153,187]. A calculated spin splitting is interpreted together with magnetic stability, the physical origin of the splitting, and the symmetry retained under realistic surfaces and substrates.

For an M1 or M2 entry, thickness, stacking, stoichiometry, oxidation state, and domain population are documented in the same material form used for band or transport measurements. ARPES can test the predicted momentum-dependent band splitting, whereas spin-resolved ARPES or equivalent spin-sensitive band mapping provides the most direct test of the predicted spin polarization [79,168-170,173]. Dichroic and collective-excitation probes provide complementary symmetry information, as discussed in Section 5.6. Transport provides symmetry evidence when its angular and domain dependence follows the predicted spin-polarization pattern and control measurements resolve contributions from coexisting weak moments, SOC-induced splitting, interfaces, defects, or noncollinear order [155,157-158].

A phase assignment should relate complementary measurements on the relevant material form under comparable structural and control conditions. Same-specimen or in situ measurements can strengthen this correlation where feasible. Device validation additionally requires the magnetic state and symmetry conditions to be rechecked after contacts, dielectrics, gates, and encapsulation are introduced. For devices based on domain-dependent readout, reproducible control of the domain state is also needed.

### 5.6. Experimental Diagnostics: Magnons, Domains, and Transport

Collective excitations provide a complementary test of altermagnetic symmetry. Altermagnetic exchange symmetry can lift the degeneracy of oppositely polarized magnon branches without requiring strong Dzyaloshinskii–Moriya interactions. X-ray magnetic circular dichroism in a 30 nm $\alpha$-MnTe film provides a domain-sensitive signature [221], and inelastic neutron scattering probes magnons in bulk MnTe crystals [222]. Resonant inelastic X-ray circular dichroism has mapped altermagnetic magnons in bulk CrSb [223]. More broadly, spin space group theory predicts unconventional magnon degeneracies and chiral branches in collinear magnets [224].

Orientation-sensitive thermoelectric and transport measurements provide additional diagnostics. Anomalous Nernst effect (ANE) responses in epitaxial $Mn_5Si_3$ films [225] and the reported spin-splitting magnetoresistance in single-variant $RuO_2$(101)/CoFeB bilayers [164] illustrate complementary thermoelectric and spin-transport diagnostics. Direction-dependent conduction and crystal-symmetry manipulation in CrSb, together with measurements on layered $CoNb_4Se_8$, show how crystal orientation, domain selection, and material form shape transport signals [226-228]. Form-specific measurements in monolayer or ultrathin $CoNb_4Se_8$ would extend this comparison into the M1/M2 regime. These signals support symmetry analysis when linked to the relevant magnetic structure and domain state.

Domain control is equally important. Opposite altermagnetic domains can produce signals of opposite sign, so a multidomain sample can show a reduced or averaged net response. Direct domain imaging in MnTe [171] and X-ray magnetic circular dichroism in $\alpha$-MnTe [221] establish complementary routes to domain identification, while crystal-symmetry control in CrSb demonstrates active selection of the order [227]. For a 2D material, moderate spin splitting with reproducible domain control and clean interfaces may be more useful than a larger calculated splitting in a material whose domains cannot yet be prepared reproducibly.

## 6. Device Readiness and Spintronic Opportunities

Cross-order device comparisons bring together readout, stray fields, and spin-dependent transport (Table 3). Ferromagnets offer direct moment-based readout, with magnetostatic and domain costs. Ferrimagnets span finite-moment and fully compensated states, making their readout and dynamics sensitive to sublattice balance. Conventional antiferromagnets and altermagnets both offer low-stray-field operation; altermagnets additionally permit alternating spin polarization in momentum space. Device readiness depends on how these properties translate into reproducible writing, storage, and readout under operating conditions. Figure 14 presents spin-filter, spin–orbit torque, twist-controlled tunneling, and antiferromagnetic tunnel-junction examples.

Published device values show the range of operating metrics, each under its specific measurement conditions. $CrI_3$ spin-filter junctions produce 310–530% tunneling magnetoresistance (TMR) in bilayers and approximately $1.9 \times 10^4$% in four-layer barriers at cryogenic temperature [23]. Another $CrI_3$ study found up to 10,000% TMR [24]. At 300 K, $Bi_2Te_3/Fe_3GeTe_2$ spin–orbit torque devices switch at approximately $2.2 \times 10^6$ A·cm$^{-2}$ with an in-plane assist field of ±2 kOe [229], whereas $Fe_3GaTe_2/WTe_2$ devices switch at approximately $2.23 \times 10^6$ A·cm$^{-2}$ without an applied field when current flows along the $WTe_2$ *a* axis [230]. Compensated tunnel platforms provide large readout signals in distinct regimes. CrSBr spin-filter devices show up to $4.7 \times 10^4$% TMR at 5 K, and $Mn_3Pt$-based all-antiferromagnetic tunnel junctions show approximately 100% TMR at room temperature [134,231]. The $(Fe_{0.6}Co_{0.4})_5GeTe_2$ (FCGT)/$WSe_2$/FCGT junction produces approximately 75% TMR at 10 K [232]. Comparisons should account for differences in geometry, temperature, thickness, measurement protocol, and readout definition.

Material-level benchmarks answer a different question. CrSb and MnTe combine momentum-resolved spectroscopy with domain or symmetry control across defined film and bulk forms [168-173,227]. Their band-splitting values depend on material form, momentum path, energy window, and analysis protocol; Section 5.2 gives the corresponding sample and probe context.

### 6.1. Ferromagnetic and Ferrimagnetic Devices

Ferromagnetic 2D devices are among the most mature platforms because they provide large magneto-optical, anomalous Hall, and tunneling readout signals together with established spin-torque switching routes. Vertical spin-filter junctions based on $CrI_3$ exploit layer-by-layer magnetic configurations in atomically thin tunnel barriers [23-24]. In conventional magnetic junctions, tunnel conductance depends on the relative magnetization of the electrodes [233-235]. The spin Hall effect provides a means of generating spin currents [236], and current-driven torques can switch perpendicular magnetization [237]. Metallic $Fe_3GeTe_2$ and $Fe_3GaTe_2$ provide planar electrical readout and current-driven magnetization switching [86-87,91-92,229-230,238-239]. The main constraints for dense integration are stray fields, domain formation, oxidation, interface scattering, contact resistance, and Joule heating.

Switching currents and fields depend strongly on the adjacent spin-current source, tunnel barrier, and encapsulation. MBE growth has yielded wafer-scale $Fe_3GeTe_2$ films [64] and large-area $Fe_5GeTe_2$ films [65]; CVD provides 1T-$CrTe_2$ flakes [66]. Material and device consistency therefore requires control over Fe occupancy, self-intercalation, phase purity, and oxidation [72-74]. Ferrimagnetic 2D devices remain at an early stage. $Cr_2S_3$ nanosheets and non-vdW ferrite nanosheets provide ultrathin finite-moment platforms reported as ferrimagnetic [70,97-98]. Fully compensated ferrimagnetic candidates add prospective spin-split electronic channels with zero net spin moment [101-106]. A key materials goal is to combine stable compensation, a reproducible readout, and wafer-compatible growth in one 2D platform.

### 6.2. Antiferromagnetic Devices

Ideally compensated antiferromagnetic states can combine low stray fields with exchange-scale dynamics and symmetry-based functionality, although practical readout requires state-sensitive probes. Spin-filter tunneling in CrSBr [134] and all-antiferromagnetic tunnel junctions [231-232] provide electrical readout routes. The optical and local magnetic probes discussed in Sections 4.1–4.2 offer complementary sensitivity to magnetic symmetry and domain structure. Nonvolatile electrical control has also been demonstrated in CrSBr [138]. $MnBi_2Te_4$ films add topological transport functionality, with observed device behavior sensitive to thickness, disorder, and Fermi-level position [113,144-145]. Operating temperatures remain below room temperature in the $MPX_3$ and CrSBr examples discussed here. Environmental stability and tunnel-barrier functionality remain material specific.

For antiferromagnetic devices, deterministic writing and reading of a domain or Néel-vector state are central readiness metrics alongside ordering temperature. $Mn_3Pt$/MgO/$Mn_3Pt$ junctions with an MnPt pinning layer [231] and FCGT/$WSe_2$/FCGT junctions [232] demonstrate compensated tunnel readout in distinct interface and symmetry settings. Other experimental routes include spin-phase-change memory [240], octupole-driven $Mn_3Sn$ junctions [241], twisted CrSBr junctions [242], and spin detection using an asymmetric $Mn_3Sn$/CoFeB junction [243]. In antiferromagnetic tunnel junctions, momentum-dependent spin polarization and Néel spin currents can enable magnetic-state readout, while spin torques offer electrical control of the Néel vector [244]. Calculations predict nearly complete spin polarization at selected momenta in $Mn_3GaN$ [245]. For comparison, calculations predict multistate operation in ferromagnetic $Fe_3GaTe_2$/$\alpha$-$In_2Se_3$/$Fe_3GaTe_2$ junctions [246]. Twisted CrSBr experiments demonstrate nonvolatile states in both mixed monolayer/bilayer stacks and all-antiferromagnetic stacks of three bilayers [247]. These examples connect symmetry-based readout to the wider spin-transport, spin-torque, and Néel-vector-control framework [248-252]. Fast intrinsic dynamics also motivate terahertz and magnonic applications in vdW heterostructures [253-255].

### 6.3. Prospective Altermagnetic Devices

At the diagnostic stage, Hall bars, anisotropic-magnetoresistance structures, tunnel junctions, optical-dichroism devices, and terahertz test structures can establish whether a symmetry- and domain-dependent response tracks a well-characterized magnetic state when control experiments distinguish competing mechanisms [256].

At the control stage, devices that encode information in magnetic domains or the Néel vector should be evaluated for writing reproducibility, operating temperature, switching current or field, retention, endurance, and interface sensitivity. Domain imaging and control in MnTe provide thin-film benchmarks [171]. CrSb film heterostructures show room-temperature anomalous Hall readout and current-driven Hall-resistance switching [227].

At the functional stage, a controlled state supports reproducible symmetry-dependent spin polarization, switching, and device-compatible readout in a defined M1 or M2 form. Spin splitters, tunnel memories, nonreciprocal elements, and terahertz oscillators could combine antiferromagnetic compensation with ferromagnet-like spin-split bands [34-36,151-152,256]. AHE, ANE, magnetoresistance, and terahertz responses provide supporting diagnostics when their symmetry or domain dependence is demonstrated and controls exclude competing mechanisms [158,167,225,257]. Topological Hall-like signals call for a separate assessment of noncoplanar spin textures and interfacial origins [258-259].

### 6.4. Ferroic Control and Multiferroic Heterostructures

Ferroic order and heterostructure design provide a route to nonvolatile symmetry control. Single-layer multiferroicity in $NiI_2$ [260] and spin-induced ferroelectricity in bulk $CuCrP_2S_6$ [261] illustrate coupled electric and magnetic order. Broader multiferroic and ferroelastic studies provide polarization- and strain-based design routes [262-266]. The theoretical routes in Section 5.4 use external electric fields, polar distortions, strain, or sliding to control spin splitting [207-216]; hidden sector-resolved polarization provides a distinct possibility in globally spin-degenerate antiferromagnets [217].

A practical heterostructure strategy is to assign writing to a spin–orbit or ferroic control layer, state storage and spin splitting to the compensated magnetic layer, readout to a tunnel electrode or spin-selective channel, and environmental protection to an encapsulating layer. Success depends on preserving phase purity, stacking, domain state, and interfacial symmetry after assembly. Near-term work should establish reproducible structures that separately validate readout, switching, and interface stability before these functions are combined in integrated devices.

## 7. Conclusions and Outlook

Two-dimensional magnetism is moving from thickness-dependent phase discovery to magnetic-symmetry engineering. The comparison reveals distinct maturity profiles: finite-moment magnets currently lead in direct readout and device integration; compensated antiferromagnets offer low-stray-field operation and exchange-scale dynamics; and altermagnets add momentum-dependent spin polarization, with current experimental support centered on M2 thin films and M3 bulk benchmarks. The next major goal is a fully verified atomically thin altermagnetic phase: an M1 material in which magnetic structure, symmetry, and spin polarization are established in the same form under matched conditions.

### 7.1. Materials and Device Readiness

The more mature device platforms considered here include finite-moment spin filters, magnetic tunnel junctions, spin–orbit torque devices, and antiferromagnetic tunnel junctions; opto-spintronic structures provide established diagnostic platforms. Ferromagnets generally offer stronger direct readout, ideally compensated antiferromagnetic states can minimize stray fields, and ferrimagnets span tunable finite-moment and fully compensated regimes. For devices that encode information in magnetic domains or the Néel vector, reproducible state preparation is essential. For all classes, thickness uniformity, environmental stability, phase purity, domain structure, and retention after gating, encapsulation, contact fabrication, and thermal cycling are as important as nominal Curie or Néel temperatures.

In an integrated device, the central question is whether the measured response follows the intended magnetic structure and domain state under operating conditions. The phase-level criteria in Section 5.5 and the device metrics in Section 6.3 provide complementary tests of this connection.

### 7.2. Experimental Priorities and Integration Challenges

Atomically thin altermagnetism will require a growth or thinning route that preserves the requisite magnetic symmetry. Epitaxial film studies [168,185], calculations for potentially exfoliable $RuF_4$ and $VX_4$ monolayers [157,198], and twisted- or stacked-bilayer models [200-202] offer distinct starting points. Thickness-dependent structural, magnetic, and band measurements would establish where the target order and spin texture survive, and how substrates and encapsulation alter that window.

Where spin-resolved ARPES is impractical, complementary probes can narrow the magnetic assignment. Local magnetic imaging [78,171] and diffraction [175,186] constrain the order; dichroism [221], spin-integrated ARPES [168,173], and magnon spectroscopy [222-223] test its spatial, electronic, and collective signatures. A useful next step is to correlate these measurements across a controlled thickness series and determine which combinations remain sensitive in device-sized samples.

Reproducible domain writing is the next device-level target. MnTe domain imaging [171] and crystal-symmetry control in CrSb [227] provide experimental starting points. The priority now is repeated write-read cycling in thin devices, with switching thresholds, retention, and endurance measured for the same domain configuration.

Integration studies should track the magnetic response through contact deposition, gating, and encapsulation. Dry-transfer assembly [58-59], tunnel junctions [24,231-232], and spin–orbit torque devices [229-230] provide fabrication approaches to build on. Comparing magnetic and electronic signatures before and after each processing step would help identify compatible materials for writing, readout, and protection.

**Acknowledgements**

Z.M. acknowledges financial support from the National Natural Science Foundation of China (no. 524B2003). Z.L. acknowledges financial support from the Beijing Natural Science Foundation (no. JQ23005). P.Q. acknowledges financial support from the National Natural Science Foundation of China (nos. 52671300 and 52401300)). L.L. acknowledges financial support from the National Natural Science Foundation of China (no. 525B2008). Z.L. acknowledges financial support from the National Natural Science Foundation of China (nos. 52425106, 52121001, and 52271235). Z.L. acknowledges financial support from the National Key R&D Program of China (no. 2022YFA1602700). X.W. acknowledges financial support from the National Natural Science Foundation of China (no. 52601331). This work was supported by the National Natural Science Foundation of China (no. U25A20244). This work was supported by the Fundamental Research Funds for the Central Universities.

**Conflict of Interest**

The authors declare no conflict of interest.

**Data Availability Statement**

The data that support the findings of this study are available from the corresponding author upon reasonable request.

**Table 1. Criteria for distinguishing altermagnetism from conventional antiferromagnetism and related spin-splitting phenomena.**

| Criterion | Key distinction |
|---|---|
| Organizing coordinates and defining test | Organize by net magnetization and spin-polarization patterns [34-36,147-152]. Assign magnetic order from structure and full symmetry [81-83,154]; in 2D, examine all spin layer operations protecting nonrelativistic degeneracy [153]. |
| Distinguish from neighboring cases | Fully compensated FiM: possible *s*-wave splitting [101-106].<br>AM: SOC-induced weak moments can coexist [155,157].<br>Rashba/Dresselhaus: SOC-driven splitting [36].<br>Berry-curvature and multipolar responses can occur in both collinear AM and noncollinear AFM [149,158]. |
| Evidence package | Match magnetic structure and symmetry [159] with domain orientation [171], band dispersion [168,170,173], and, where feasible, directly measured spin polarization [169] in the same material form. Correlated measurements provide the strongest support; complementary probes include transport [36], dichroism [221], and magnons [222-224]. |

**Table 2. Reporting dimensions and minimum requirements for two-dimensional magnetic phase assignments.** The M1–M4 reporting classes describe material form and realization status. The evidence category is reported separately as direct measurement, symmetry-sensitive or supporting measurement, indirect inference, or calculation, and ordering-temperature provenance is identified as measured, inferred, or calculated.

| Reporting dimension | Information to report |
|---|---|
| Material form | Layer count/thickness; phase purity, structure and stacking; growth route; substrate; control conditions; composition; oxidation state and stability [37-38,41]. Use few-layer for discrete layer counts. |
| Magnetic order | State the evidence basis: direct magnetic-order measurement; symmetry-sensitive/supporting measurement; indirect inference; or calculation. Form-appropriate probes include magnetometry/imaging [78,123,133], magneto-optics [11-13], electrical transport [14,122], Raman and optical symmetry probes [75-77,107,111,118-119], tunneling [134], neutron scattering [114-117], and $\mu$SR [80,132]. |
| 2D altermagnetism | Document full spin layer symmetry and its SOC-dependent counterpart [36,153,256]. Relate structure and domains to band measurements [79,168-171,173]; dichroism and magnons provide complementary tests [221-224] under comparable conditions. Assess relevant competing contributions from weak moments [155,157], SOC and impurities [257], noncollinear order [158,258], and interfaces [259] using appropriate controls [167]. |
| Reporting class | M1: Realized, explicitly layer-resolved atomically thin form.<br>M2: Thickness-defined thin flake, film, or nanosheet without an established discrete layer count.<br>M3: Bulk or thick-film benchmark without a demonstrated thickness-dependent 2D state.<br>M4: Theoretical candidate, calculation, design route, or screening framework.<br>Explain borderline M2/M3 assignments; evidence category and temperature provenance remain independent. |

**Table 3. Qualitative device-readiness considerations for two-dimensional magnetic orders**. Representative quantitative anchors and supporting literature are discussed in Section 6.

| Reporting criterion | FM | FiM | AFM | Prospective altermagnets |
|---|---|---|---|---|
| Operating temperature | Material dependent: selected devices switch at room temperature; many operate below it | No general benchmark; sparse 2D device data | Most demonstrated 2D devices operate below room temperature; topological devices often cryogenic | Device-specific; M1 operating-temperature benchmarks unestablished |
| Readout mechanisms | Large MOKE, RMCD, AHE, and TMR signals | Moment-based away from compensation; spin-split readout proposed for fully compensated states | Indirect optical, tunneling, SHG, LD, or topological readout | Symmetry-sensitive Hall and optical signals; M1 spin-split readout remains prospective |
| Switching | Field, gating, and spin–orbit torque in selected systems | Compensation may enable fast dynamics; suitable 2D platforms still needed | Deterministic domain or Néel-vector writing | Current-driven magnetic-state control in selected films; M1 switching unestablished |
| Domain control | Domain variability can limit reproducibility | Sublattice balance and domain selection both matter | Domain selection often harder than order detection | Domain selection for reproducible state-dependent output |
| Air stability | Many Cr halides and some Fe tellurides require encapsulation | Material dependent; oxide and nanosheet routes differ | CrSBr is relatively stable; many systems are sensitive | Unknown for most M4 2D candidates |
| Wafer-scale compatibility | Wafer-scale MBE demonstrated for $Fe_3GeTe_2$ | At an early stage | Growth and assembly are emerging | Mostly prospective |
| Interface sensitivity | Oxidation, contact resistance, Joule heating | Strongly composition dependent | Interface symmetry controls tunnel readout | Preserve symmetry required for spin splitting |
| Integration difficulty | Lower in selected mature systems with direct readout and switching | High: few established 2D materials | High: complex writing and readout | High: magnetic order, spin splitting, and domain state require verification in the device |

**Table 4. Representative ferromagnetic, ferrimagnetic, and complex multisublattice two-dimensional materials and layered benchmarks.**

| Material / system | Magnetic order | Material form | Temperature / provenance | Magnetic evidence / probe | M class | Additional notes | Refs. |
|---|---|---|---|---|---|---|---|
| **Insulating / semiconducting ferromagnets** | | | | | | | |
| $CrI_3$ | Ising FM insulator; layer-dependent interlayer order | Exfoliated 1L and few-layer vdW flakes | Measured: 1L $T_C$ = 45 K; bulk $T_C$ = 61 K | Direct: 1L Kerr microscopy | M1 (1L/few-layer) M3 (bulk) | Bilayer: suppressed net moment; metamagnetism. | [11] |
| $CrBr_3$ | FM semiconductor | Exfoliated 1L flakes | Measured: 1L $T_C$ = 34 K; bulk $T_C$ = 37 K | Supporting: 1L magnetophotoluminescence | M1 (1L) M3 (bulk) | Optical signature of spontaneous magnetization. | [54] |
| $CrCl_3$ | Easy-plane 2D-XY FM monolayer | MBE-grown 1L on graphene/6H-SiC (0001) | Measured: 1L $T_C$ = 12.95 ± 0.03 K | Direct: 1L in situ XMCD; critical scaling | M1 | BKT transition not established. | [53] |
| $Cr_2Ge_2Te_6$ | Soft FM semiconductor | Exfoliated bilayer to bulk flakes; small perpendicular field in key measurements | Measured: bilayer $T_C^*$ ~30 K at 0.075 T; bulk-limit $T_C^*$ ~68 K under field; bulk $T_C$ near 61 K | Direct: magneto-optics under field | M1 (2L; field-defined) M3 (bulk) | No zero-field monolayer $T_C$ assigned. | [12] |
| $CrSiTe_3$ | FM semiconductor | Exfoliated flakes: 8.5, 20, 36, and 125 nm (resistivity); monolayer model | Inferred: thin-flake resistivity anomaly 80–120 K; Calculated: 1L $T_C$ ~80 K; Measured: bulk $T_C$ = 33 K | Indirect: thin-flake resistivity; Calculated: monolayer Ising-model estimate | M2 (flakes) M3 (bulk) M4 (candidate; 1L theory) | 7 nm flake: p-type field effect down to 6 K. | [55] |
| **Metallic ferromagnets** | | | | | | | |
| $Fe_3GeTe_2$ | Itinerant FM metal with PMA | Exfoliated 1L to bulk-like flakes | Measured: $T_C$ ~207 K (bulk-like); ~130 K (1L) | Direct: 1L RMCD and MOKE | M1 (1L/few-layer) M3 (bulk-like) | 1L scaling consistent with 2D Ising behavior. | [13] |
| $Fe_3GeTe_2$ | Itinerant FM metal | Wafer-scale 8 nm MBE film | AHE-derived (Arrott fit): $T_C$ = 216.4 K | Direct: film SQUID magnetometry; Supporting: AHE | M2 | Composition- and interface-dependent $T_C$. | [64] |
| $Fe_3GeTe_2$ | Gate-enhanced FM metal | Ionic-gated trilayer and four-layer devices | AHE-extrapolated (0.01 T; gated trilayer): $T_C$ reaches 300 K | Indirect: gate-dependent AHE | M1 (gated) | Li-ion intercalation under gating. | [14] |
| $Fe_{5-x}GeTe_2$ | Near-room-temperature metallic FM | Cleavable bulk crystals; flakes: 10–30 nm (RMCD), 12–150 nm (transport) | Measured: bulk $T_C$ ~310 K (thermally cycled); flakes ~280 K (AHE-derived) | Bulk magnetometry; flake RMCD and AHE | M2 (flakes) M3 (bulk) | Nominal $Fe_5GeTe_2$; $T_C$ depends on stoichiometry and thermal history. | [88] |
| $Fe_5GeTe_2$ | Itinerant FM metal | MBE-grown 12 nm film and bilayer on insulating substrate | Measured: 12 nm film $T_C$ ~293 K; bilayer $T_C$ ~229 K | Direct: SQUID (bilayer and 12 nm); XMCD (12 nm). | M1 (bilayer) M2 (12 nm) | Layer-by-layer growth; nominal 5:1:2 composition confirmed at 12 nm. | [65] |
| $Fe_3GaTe_2$ | Above-room-temperature FM metal with large PMA | 2.7 nm (3-layer) nanosheet: structure only; 9.5–112 nm nanosheets: transport; bulk crystal | Measured: $T_C$ ~350–380 K | Mixed: AHE (supporting), MFM (direct), 9.5–112 nm | M1 (3-layer structure); M2 (9.5–112 nm magnetic evidence); M3 (bulk) | 3-layer evidence: structural only. | [89] |
| $CrTe_2$ | Metallic FM; phase- and thickness-sensitive | MBE-grown single- and few-layer films on bilayer graphene | Measured: few-layer $T_C$ up to ~300 K; 1L ~200 K | Direct: SQUID, XMCD; Supporting: ARPES | M1 | Phase-, substrate-, and thickness-dependent assignment. | [90] |

| Material / system | Magnetic order | Material form | Temperature / provenance | Magnetic evidence / probe | M class | Additional notes | Refs. |
|---|---|---|---|---|---|---|---|
| $CrTe_2$ | Metallic FM; polymorph sensitive | CVD 1T flakes: 7.6–130 nm ($T_C$ series); ~1.2 nm bilayer: structural characterization | Measured (out-of-plane RMCD/AHE): $T_C$ ~167–212 K for CVD 1T flakes | Direct/supporting: VSM, SQUID, RMCD, transport | M1 (bilayer structure); M2 (7.6–130 nm magnetic evidence) | Bilayer evidence: structural only. | [66] |
| **Ferrimagnetic or complex multisublattice systems** | | | | | | | |
| $Cr_2S_3$ | Reported ferrimagnetic semiconductor | One-unit-cell CVD nanosheets; non-vdW ultrathin evidence | Measured response: reported $T_N$ up to 120 K | Finite-moment evidence: magnetization | M1 | Antiparallel sublattices unresolved. | [97] |
| $CoFe_2O_4$ | Reported ferrimagnetic oxide | Nanosheets down to one unit cell (~1.7 nm); magnetic measurements on thicker nanosheets and ensembles | Not assigned: no transferable $T_C$ | Ensemble VSM; thickness-resolved MFM and MOKE on thicker nanosheets | M1 (one-unit-cell structure); M2 (thickness-defined magnetic nanosheets) | One-unit-cell order and antiparallel sublattices unconfirmed. | [70] |
| $Mn_3Si_2Te_6$ | Ferrimagnetic layered semiconductor with competing exchange | Single crystals; layered bulk benchmark | Measured: $T_C$ ~78 K | Direct: bulk magnetometry and magnetic-structure study | M3 | Monolayer order untested. | [99-100] |

Ordering temperatures depend on dimensionality, synthesis route, stoichiometry, field, and device configuration. Reporting classes describe the material form and control condition in each row. Transport-derived temperatures specify the extraction criterion. The evidence column separately describes how magnetic order was identified. $T_C^*$ denotes a field-defined ferromagnetic-like transition temperature. 1L, monolayer; AHE, anomalous Hall effect; BKT, Berezinskii–Kosterlitz–Thouless; CVD, chemical vapor deposition; M1, explicitly layer-resolved atomically thin material form in the stated control condition; M2, thin-flake, film, or nanosheet realization without an explicitly established atomically thin layer count; M3, bulk or thick-film benchmark; M4, theory or high-throughput candidate; FiM, ferrimagnetic; MBE, molecular beam epitaxy; MFM, magnetic force microscopy; MOKE, magneto-optical Kerr effect; PMA, perpendicular magnetic anisotropy; RMCD, reflective magnetic circular dichroism; VSM, vibrating-sample magnetometry; XMCD, X-ray magnetic circular dichroism.

**Table 5. Representative antiferromagnetic two-dimensional materials and layered benchmarks.** Layer-dependent FM counterparts are included for comparison.

| Material / system | Magnetic order | Material form | Temperature / provenance | Magnetic evidence / probe | M class | Additional notes | Refs. |
|---|---|---|---|---|---|---|---|
| $CrCl_3$ | Layered AFM with FM intralayer correlations | Bulk cleavable crystals; exfoliable vdW parent | Measured: 17 K: FM correlations or FM-like behavior; 14 K: long-range AFM | Direct: bulk magnetometry | M3 | Weak bulk interlayer AFM coupling. | [51] |
| $FePS_3$ | Ising-type zigzag AFM | Bulk to monolayer | Raman-inferred: $T_N$ ~118 K (bulk–1L) [107]. LD-inferred: $T_N$ ~117/116/97 K (3L/2L/1L) [119]. | Indirect/supporting: Raman zone folding; linear dichroism (LD) | M1 (1L) M3 (bulk) | Bulk Raman signatures corroborated by susceptibility. | [107,119] |
| $MnPS_3$ | Heisenberg-type Néel AFM | Bulk to bilayer; monolayer tested by SHG | Measured: bulk $T_N$ ~78 K; SHG-inferred: order persists to 2L; suppressed in 1L | Direct: bulk neutron [114]; Symmetry/supporting: thin-layer SHG [77] and Raman [75] | M1 (1L–few-layer); M3 (bulk) | SHG: suppressed 1L Néel order. | [75,77, 114] |
| $NiPS_3$ | XXZ- or XY-like zigzag AFM | Bulk and 1L–few-layer flakes | Measured: bulk $T_N$ ~155 K (susceptibility), 158.65(2) K (neutron fit); Inferred: order persists to 2L; no 1L ordering signature down to 25 K | Direct: bulk neutron [115]; Indirect/supporting: few-layer Raman [118] | M1 (1L–few-layer); M3 (bulk) | Slight $T_N$ reduction toward 2L. | [115,118] |
| CrSBr | Bulk/few-layer A-type AFM; strong intralayer FM | Bulk crystals; 2–6-layer vdW flakes and devices | Measured: bulk $T_N \approx$ 131–132 K; symmetry-sensitive SHG: 2L $T_N$ = 140.0 ± 0.1 K and 6L $T_N$ = 137.9 ± 5.2 K; transport-inferred: 2–6L transitions, including ≈ 135 K for 4L. | Direct, bulk: magnetometry [109,123]. Symmetry-sensitive, bilayer/multilayer: SHG [111]. Supporting, bilayer/multilayer: magnetotransport [122]. | M1 (2–6L); M3 (bulk) | Even layers: compensated; odd layers: net layer moment. | [109,111, 122-123] |
| CrSBr | Experimental monolayer FM semi-conductor | Exfoliated isolated monolayer | Inferred: 1L $T_C$ = 146 ± 2 K | Symmetry-sensitive: SHG | M1 | Magnetic-dipole SHG supports inversion-symmetric FM. | [111] |
| CrOCl | Low-symmetry vdW AFM insulator | Bulk crystals; exfoliated ~22 nm flakes | Measured: bulk $T_N$ ~13.6 K; anomalies 27.3/28.2 K; ~22 nm flakes: Raman anomaly ~27 K, $T_N$ unassigned | Measured: bulk magnetization/heat capacity; Indirect: flake Raman | M2 (~22 nm) M3 (bulk) | Bulk 27.3/28.2 K anomalies unresolved. | [67] |
| $MnBi_2Te_4$ | A-type magnetic topological material | Bulk crystals and exfoliated septuple-layer films | Measured: bulk $T_N$ = 24.2(5) K (magnetometry) [112]; Transport-inferred: ≈ 18/21/23 K (3/4/5 SL) [113] | Direct: bulk magnetometry [112]; Indirect: 3–5 SL resistance anomalies [113] | M1 (3–5 SL) M3 (bulk) | Topology depends on parity, field, gating, and disorder. | [112-113, 144-145] |
| $MnBi_2Te_4$ | Calculated thickness-dependent magnetic order | 1-SL and 2-SL theoretical Monte Carlo models | Calculated: 1-SL FM $T_C$ ~12 K; 2-SL AFM $T_N$ ~24.4 K | Calculation: Monte Carlo | M4 (calc.) | Ordered model; Mn/Bi intermixing excluded. | [142] |

Rows distinguish material form/control conditions, temperature provenance, and magnetic evidence. CrSBr layer-defined states and $MnBi_2Te_4$ experimental/theoretical entries are listed separately. 1L, monolayer; 2L, bilayer; 3L, trilayer; AFM, antiferromagnetic; M1, explicitly layer-resolved atomically thin material form in the stated control condition; M2, thin-flake or film realization without an explicitly established atomically thin layer count; M3, bulk or thick-film benchmark; M4, theory; FM, ferromagnetic; SHG, second-harmonic generation; SL, septuple layer; vdW, van der Waals.

**Table 6. Experimental benchmarks, candidates, and diagnostic cases relevant to altermagnetism.**

| Material | Status / order | Material form / M class | Ordering temperature / provenance | Key evidence | 2D relevance / caveat | Refs. |
|---|---|---|---|---|---|---|
| MnTe | Benchmark altermagnet | M2: 30 nm epitaxial films. M3: 200 nm epitaxial film; bulk single crystals. Non-vdW. | M2, 30 nm: reported $T_N$ ~300 K [171]. M3 bulk: measured $T_N$ = 310 K [172]; literature 307 K [173]. 200 nm film: $T_N$ unassigned. | M2, 30 nm films: domain imaging/control [171]; XMCD [221]. M3, 200 nm film: SX-ARPES [172]. M3 bulk: ARPES [173]; SARPES [172]; INS magnons [222]. | Film/bulk benchmarks; atomically thin AM not established by cited studies. | [171-173,221-222] |
| CrSb | High-temperature metallic AM benchmark | M2: 30 nm epitaxial films. M3: bulk single crystals. Non-vdW. | M3 bulk: reported $T_N$ ~700–705 K [169-170]. M2 film: $T_N$ unassigned. | M2 film: spin-integrated SX-ARPES, ~0.6 eV [168]; symmetry manipulation [227]. M3 bulk: 3D ARPES/SARPES, ~1.0 eV and spin-polarization contrast [169]; ARPES, ~0.93 eV [170]; RIXS magnons [223]; direction-dependent conduction [226]. | Film splitting and bulk polarization: separate evidence sets; atomically thin phase unverified. | [168-170, 223, 226-227] |
| $KV_2Se_2O$ | Contested *d*-wave assignment; reported G-type bulk order | M3: layered bulk single crystals; polycrystalline powder | Measured: powder $T_N \approx 400$ K (neutron) [175]; single-crystal $T_N$ imprecise | Initial magnetic characterization and SARPES: *d*-wave assignment [174]. Later neutron diffraction: G-type AFM [175]. | Sample-specific structure–spectroscopy conflict; reduced-thickness order untested. | [174-175] |
| $Rb_{1-\delta}V_2Te_2O$ | Reported G-type bulk AFM; electronic assignment under reassessment | M3. Layered single crystals and powder | Measured: ~307 K in single crystals; 337 K in powder | Single crystals: susceptibility, spin-resolved photoemission, quasiparticle interference [176]. Powder neutron diffraction: G-type AFM [177]. | Different samples and compositions; atomically thin phase unverified. | [176-177] |
| $Cs_{1-\delta}V_2Te_2O$ | Reported hidden, layer-local altermagnetism | M3. Bulk single crystals ($\delta \approx 0.23$); preprint | Measured: $T_N \approx 257$ K | Neutron: G-type structure. Surface-sensitive SARPES: near-surface spin polarization. Calculations: layer-local spin texture [179]. | *PT*-symmetric, globally spin-degenerate bulk bands. $CsV_2Te_2O$: strain-induced moment calculated for C-, not G-type order [180]. | [178-180] |
| $CsCr_2S_2O$ | C-type AFM; calculated *d*-wave spin splitting | M3. Synthesized bulk crystals and powders | Measured: $T_N = 326$ K; metal-to-insulator transition ~305 K | Powder neutron diffraction: C-type order; transport: metal–insulator transition; spin splitting: calculated | Electronic spin polarization and atomically thin phase unverified. | [181-182] |
| $La_2O_3Mn_2Se_2$ | Layered correlated *d*-wave altermagnet candidate | M3: layered bulk material; magnetic measurements on powders | Measured: $T_N$ ~166 K; two-dimensional short-range order persists above $T_N$ | Order/correlations: magnetometry; neutron pair distribution. Magnons: powder INS; linear spin-wave model. Symmetry: material-specific *d*-wave analysis. Bands: calculated splitting; spin polarization unmeasured. | Layered bulk benchmark; atomically thin form unverified. | [183-184] |
| GdAlSi | M1 boundary case with calculated NRSS | M1: one-unit-cell and discrete few-layer epitaxial films on Si. M3: 134-ML bulk-like film; bulk crystals. Layer-resolved intermediate films form a thickness series. | M1, one-unit-cell: no direct ordering temperature. M3, 134-ML film/bulk crystal: measured $T_N \approx 32$ K. | Films: spontaneous ultrathin AHE; no discernible SQUID net moment; one-unit-cell calculation: NRSS persists after loss of the defining fourfold AM operation [185]. Bulk: polarization-resolved REXS finds cycloidal order [186]. | One-unit-cell magnetic structure and spin polarization unmeasured; outside AM class in cited model. | [185-186] |

| Material | Status / order | Material form / M class | Ordering temperature / provenance | Key evidence | 2D relevance / caveat | Refs. |
|---|---|---|---|---|---|---|
| $RuO_2$ | Contested diagnostic case | M2: 3–34 nm epitaxial $RuO_2$ films across the cited studies; $RuO_2$/permalloy bilayers contain 3–12 nm $RuO_2$. M3: bulk single crystals or powder. Rutile, non-vdW. | Unassigned: no transferable $T_N$ | Positive, probe-specific: spin-current/torque [160-162]; AHE [163]; TRSB-sensitive ARPES [79]; single-variant film transport [164]. Experimental null/conflicting: µSR and neutron order tests [80]; ARPES spin-splitting test [165]; THz IASSE test in $RuO_2$/permalloy bilayers [167]. Symmetry/diffraction reassessment [166]. | Sample-, strain-, surface-, and orientation-dependent. THz null: inverse-response test, not direct magnetic-order test. | [79-80, 160-167] |
| $CoNb_4Se_8$ | Reported layered altermagnet | M3. Layered intercalated transition-metal dichalcogenide | Measured: $T_N$ = 168 K | Structure/order: single-crystal XRD, STM, magnetometry, neutron diffraction. Band splitting: calculated. | Ultrathin magnetic phase unverified. | [228] |
| $Mn_5Si_3$ | ANE thin-film diagnostic | M2. 20 nm MBE-grown $Mn_5Si_3$(0001) epitaxial film on Si(111); non-vdW | Unassigned for the cited 20 nm film | Supporting: film ANE. Film magnetic structure and band spin polarization unmeasured. | Phase-specific structure and band tests needed. | [225] |

Key evidence is matched to material form/M class; temperature provenance is reported separately. ANE, anomalous Nernst effect; AHE, anomalous Hall effect; ARPES, angle-resolved photoemission spectroscopy; IASSE, inverse altermagnetic spin-splitting effect; INS, inelastic neutron scattering; MBE, molecular beam epitaxy; REXS, resonant elastic X-ray scattering; RIXS, resonant inelastic X-ray scattering; SARPES, spin-resolved ARPES; SQUID, superconducting quantum interference device; STM, scanning tunneling microscopy; SX-ARPES, soft-X-ray ARPES; THz, terahertz; TMD, transition-metal dichalcogenide; TRSB, time-reversal symmetry breaking; XMCD, X-ray magnetic circular dichroism; XRD, X-ray diffraction; µSR, muon spin rotation.

**Table 7. M4 candidates, design routes, and screening frameworks for two-dimensional altermagnetism.** All entries are theoretical; calculated metrics retain their model conditions. Entry types distinguish candidates, design routes, and screening frameworks. vdW, van der Waals.

| Material / route | Proposed state and form | Predicted metric | Control / design basis | Entry type | Additional notes | Refs. |
|---|---|---|---|---|---|---|
| Twisted magnetic vdW bilayers | Twisted/stacked magnetic vdW bilayers | Unassigned; material-specific $T_N/T_C$ | Stacking/registry; twist; electric-field-induced valley polarization | Design route | State/temperature depend on material and stacking; fields can alter magnetic symmetry. | [196,200-202] |
| $RuF_4$ | Predicted monolayer AM; symmetry-allowed weak moment | Calculated, no SOC: top-valence spin splitting up to 163 meV along studied *k*-path. | Material-specific calculation; weak-FM coexistence | Candidate | Predicted SOC-induced weak moment; experimental separation of signals needed. | [157] |
| $CrCl_3$ | Predicted *i*-wave AM from a FM monolayer | Calculated (GGA): conduction-band spin splitting 7.6 meV; $a/a_0$ = 0.96, $E$ = 0.40 V/Å; selected Γ–X point. | Biaxial compression, then electric-field or Janus symmetry breaking | Candidate | FM parent synthesized; proposed AM state theoretical. Related routes: $CrBr_3$, $CrI_3$. | [199] |
| $Cr_2SeO$ | Ferroic Janus $Cr_2SeO$ monolayer | Calculated: $T_N$ ~280 K; 315 K at 1% uniaxial compression | Janus polarity; uniaxial compression | Candidate | Calculated strain enhancement of $T_N$. | [208] |
| $Mn_2PSe$ | Polar, inversion-broken FeSe-type $Mn_2PSe$ monolayer | Calculated: $T_N$ ~400 K | Polar FeSe-type design; high-throughput screening | Candidate | Representative polar FeSe-type candidate. | [206] |
| AM design and magnetic-order screening | Altermagnetic and general 2D magnetic candidates | Not applicable: no single $T_N/T_C$ | Symmetry classification; AM and general magnetic-order screening | Screening | General-crystal AM screening [218]; general 2D magnetic-order screening [219]. | [71,187, 218-219] |

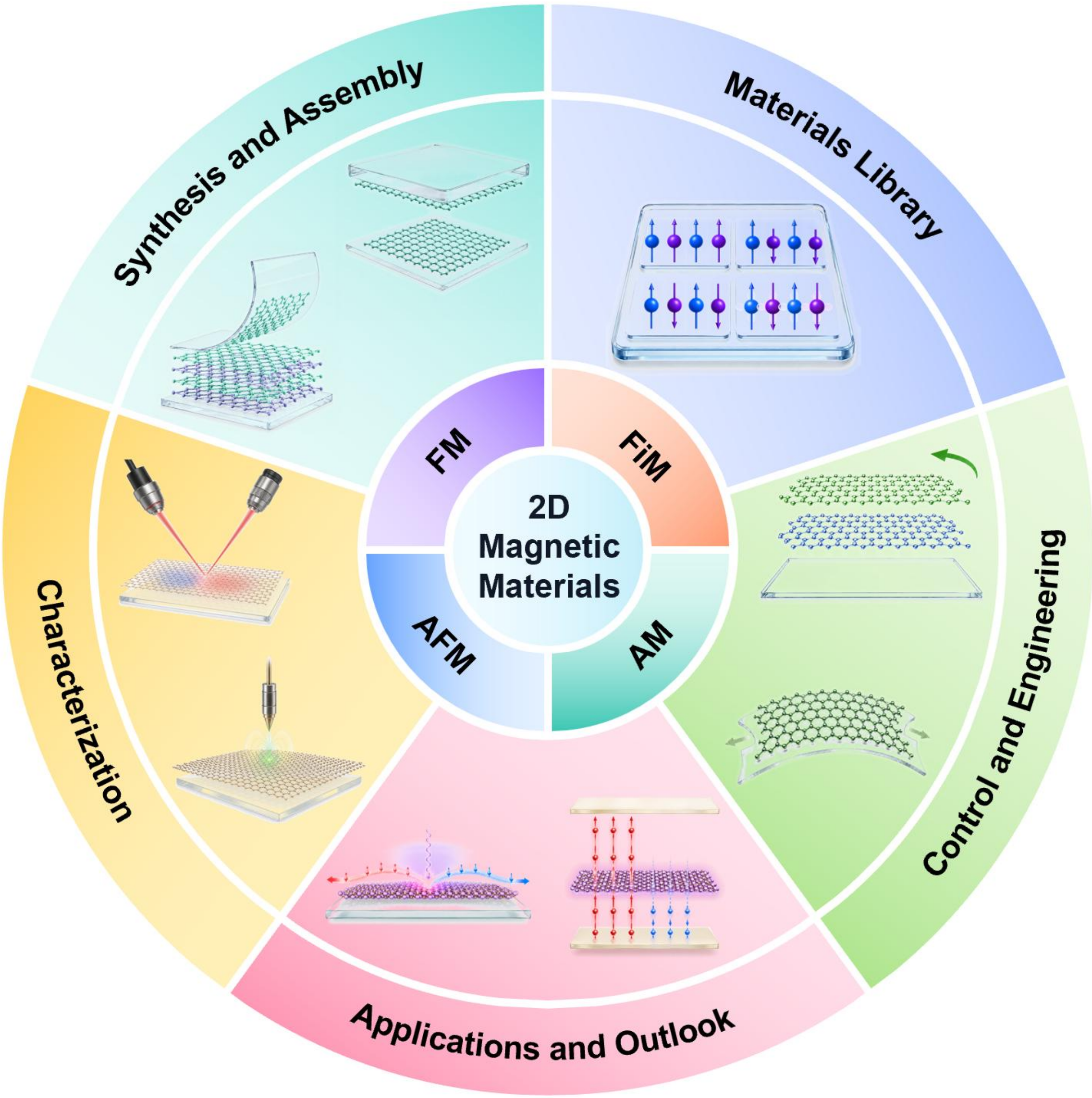

**Figure 1. Organizing framework for two-dimensional magnetic materials and their spintronic opportunities.** The radial layout links synthesis and assembly routes, representative magnetic-order families, symmetry-control variables such as stacking, twist, strain, defects, substrates, and interfaces, characterization and readout methods, and emerging device functions. The inner concentric sectors identify the four magnetic families emphasized in this Review: ferromagnets (FM), ferrimagnets (FiM), antiferromagnets (AFM), and altermagnets (AM).

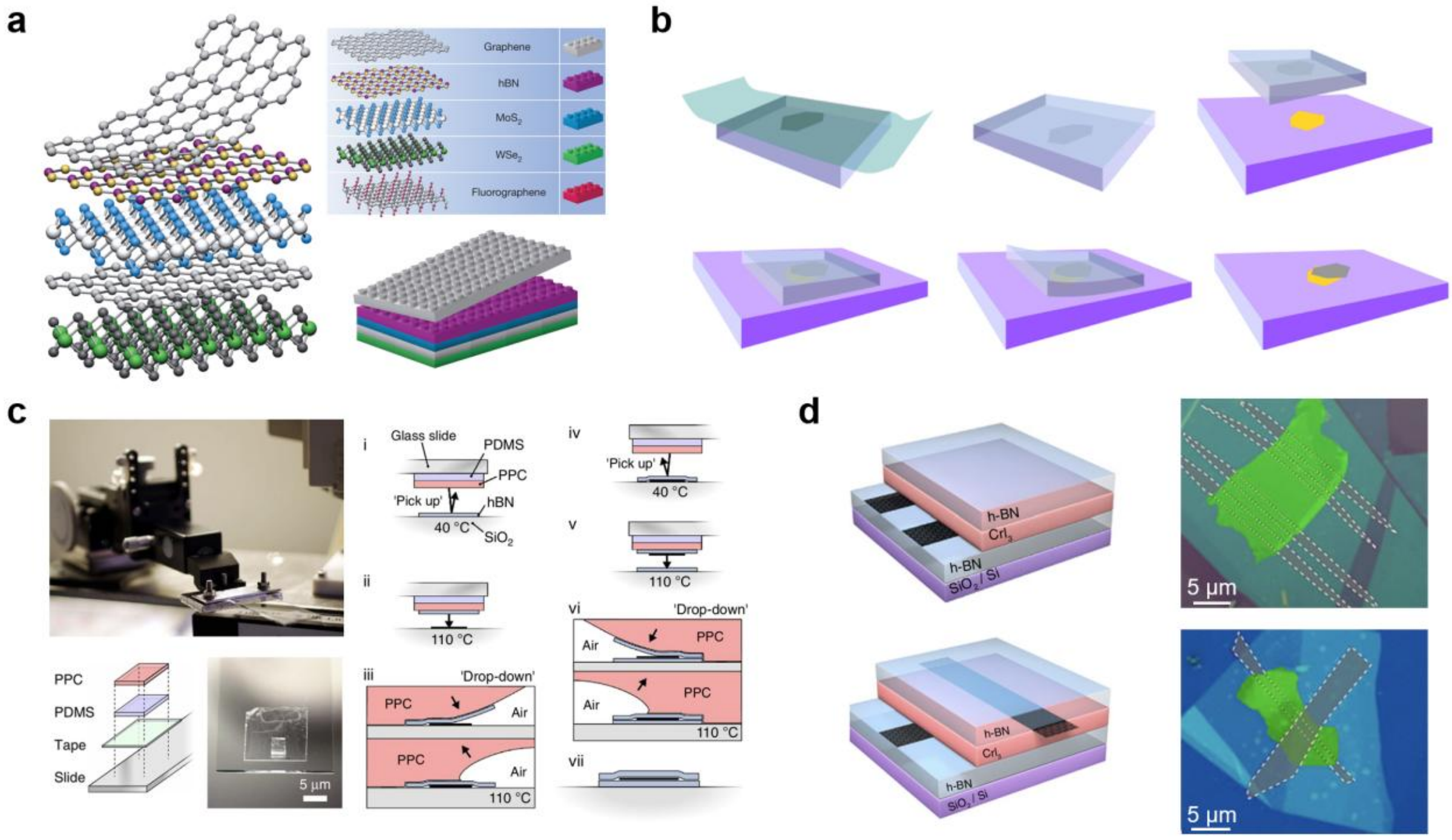

**Figure 2. Assembly routes and device architectures for van der Waals (vdW) magnetic heterostructures.** (a) The vdW Lego-block concept illustrates how atomically thin crystals can be combined into designer heterostructures [2]. Copyright 2013, Springer Nature. (b) Deterministic all-dry viscoelastic transfer: exfoliation onto the stamp, alignment with the target substrate, contact, and slow peel-off to release the flake [58]. Copyright 2014, IOP Publishing. (c) The hot pick-up technique is illustrated by the microscope-assisted transfer stage, the poly(propylene carbonate)/polydimethylsiloxane (PPC/PDMS) stamp geometry, and the temperature-controlled pickup and drop-down sequence used for polymer-assisted vdW assembly [59]. Copyright 2016, Springer Nature. (d) Hexagonal boron nitride (hBN)-encapsulated $CrI_3$ devices with graphene contacts: a lateral field-effect transistor (upper) and a vertical tunnel junction (lower), with corresponding optical micrographs [24]. Copyright 2018, Springer Nature.

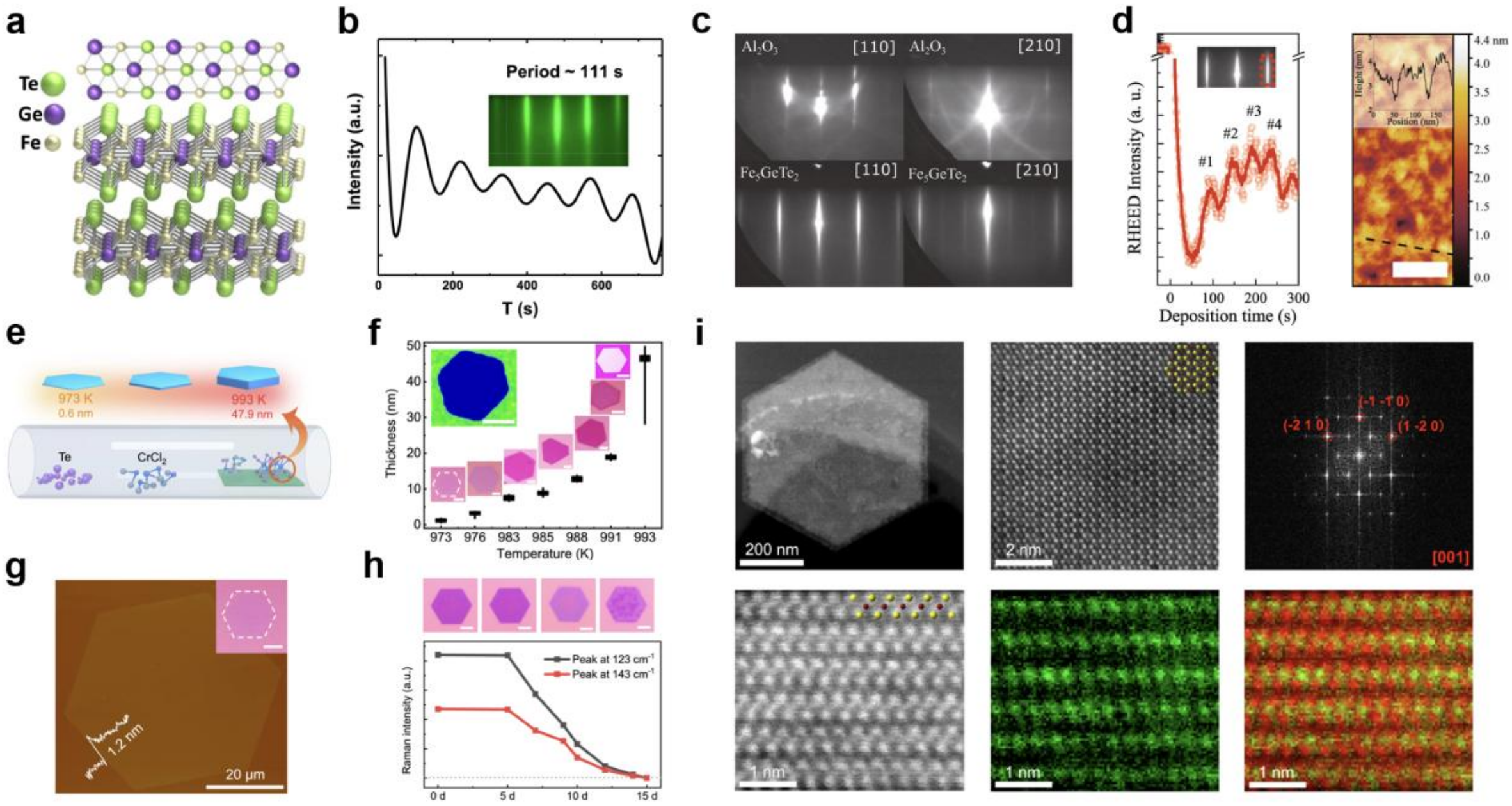

**Figure 3. Scalable growth and structural characterization of representative layered and ultrathin magnetic materials.** (a) Layered $Fe_3GeTe_2$ crystal structure, highlighting the vdW-separated slabs of this representative metallic 2D ferromagnet [64]. Copyright 2017, Springer Nature. (b) Reflection high-energy electron diffraction (RHEED) intensity oscillations during $Fe_3GeTe_2$ molecular beam epitaxy, consistent with layer-by-layer film growth [64]. Copyright 2017, Springer Nature. (c) RHEED patterns of epitaxial $Fe_5GeTe_2$ films along two substrate azimuths, showing crystallographic ordering after growth [65]. Copyright 2022, Springer Nature. (d) Growth oscillations and atomic force microscopy topography of $Fe_5GeTe_2$ films, linking deposition dynamics with nanoscale surface morphology [65]. Copyright 2022, Springer Nature. (e) Chemical vapor deposition (CVD) synthesis scheme for 1T-$CrTe_2$, including Te and $CrCl_2$ precursors, the growth substrate, and temperature-dependent crystal thickness [66]. Copyright 2021, Springer Nature. (f) Optical images showing temperature-dependent crystal thickness and morphology in CVD-grown 1T-$CrTe_2$ [66]. Copyright 2021, Springer Nature. (g) Atomic force microscopy image and height profile of a representative $CrTe_2$ flake [66]. Copyright 2021, Springer Nature. (h) Raman and optical stability measurements tracking changes in $CrTe_2$ under environmental exposure [66]. Copyright 2021, Springer Nature. (i) High-angle annular dark-field scanning transmission electron microscopy (HAADF-STEM), fast Fourier transform (FFT), cross-sectional STEM, electron energy-loss spectroscopy (EELS) mapping, and elemental overlay images showing the atomic lattice and composition of 1T-$CrTe_2$ [66]. Copyright 2021, Springer Nature.

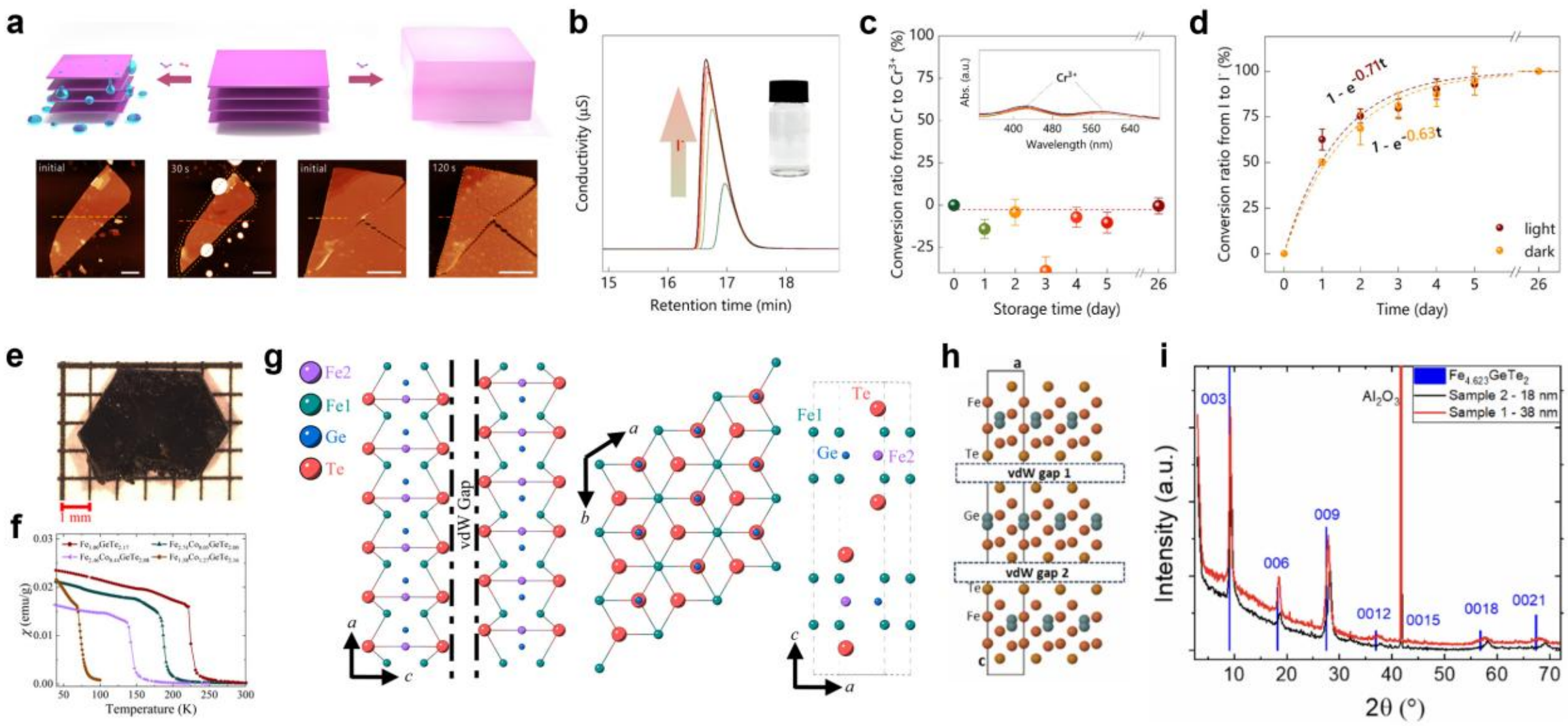

**Figure 4. Degradation chemistry, compositional control, and structural tuning in layered magnetic materials.** (a) $CrI_3$ degradation schematic and atomic force microscopy images after air exposure (30 s) or immersion in degassed water (120 s) [38]. Copyright 2022, American Chemical Society. (b) Time-dependent anion-exchange chromatograms of aqueous iodide released during $CrI_3$ hydrolysis [38]. Copyright 2022, American Chemical Society. (c) Relative aqueous $Cr^{3+}$ signal during hydrolysis, with ultraviolet–visible absorption spectra in the inset [38]. Copyright 2022, American Chemical Society. (d) Iodide-release fraction during $CrI_3$ hydrolysis under light and dark conditions [38]. Copyright 2022, American Chemical Society. (e) Unsubstituted Fe-Ge-Te single crystal grown by chemical vapor transport [73]. Copyright 2021, American Chemical Society. (f) Magnetic susceptibility of unsubstituted and Co-substituted Fe-Ge-Te crystals [73]. Copyright 2021, American Chemical Society. (g) $Fe_3GeTe_2$ structural model showing the layered arrangement and inequivalent Fe sites [73]. Copyright 2021, American Chemical Society. (h) $Fe_5GeTe_2$ crystal structure highlighting the vdW gaps [74]. Copyright 2024, American Physical Society. (i) Out-of-plane X-ray diffraction (XRD) scans of $Fe_5GeTe_2$ films with reference reflections [74]. Copyright 2024, American Physical Society.

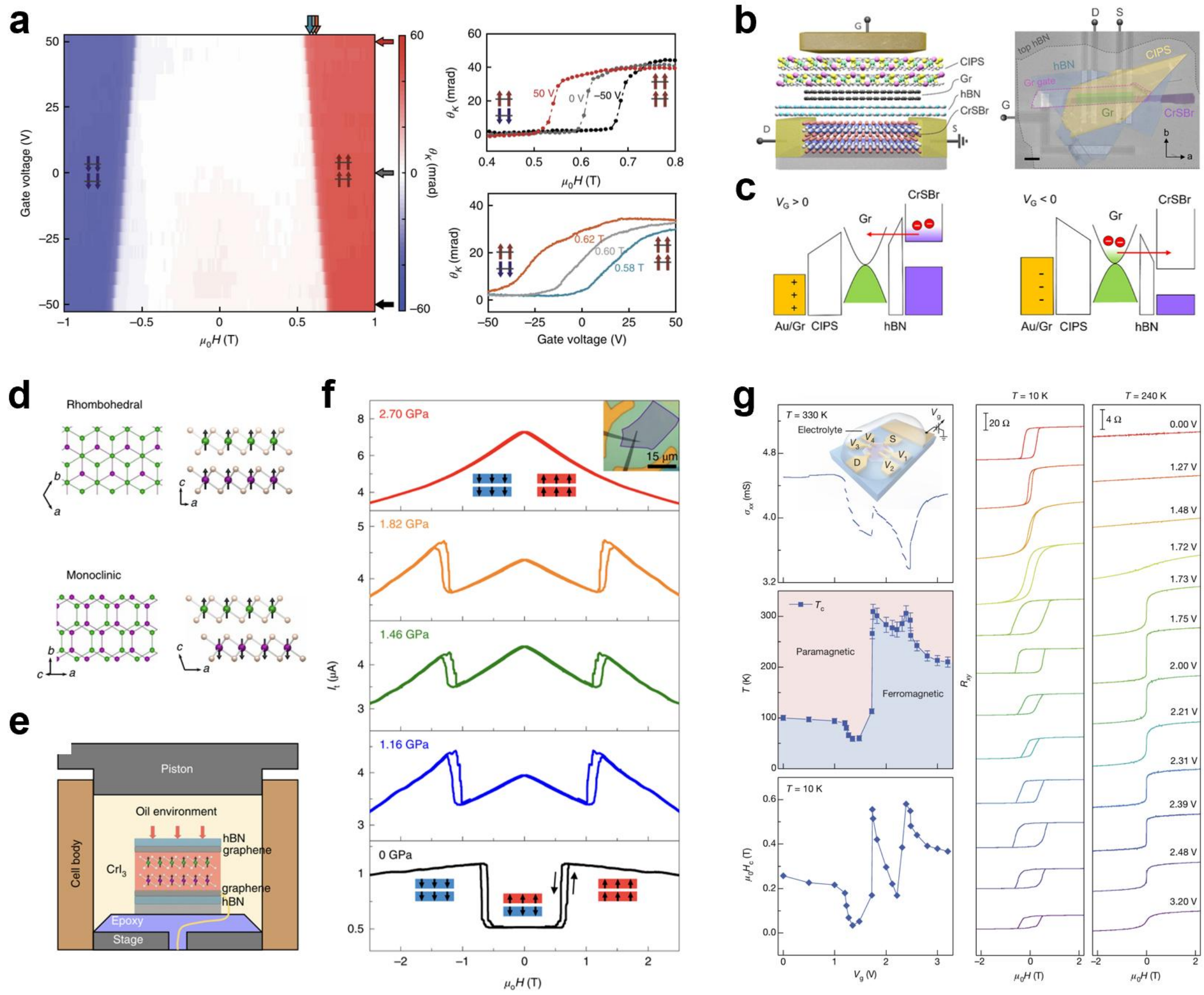

**Figure 5. External control of two-dimensional magnetism through electrostatic and ionic gating, nonvolatile floating-gate operation, and pressure.** (a) Gate-voltage- and magnetic-field-dependent magneto-optical Kerr effect (MOKE) response of bilayer $CrI_3$, including phase mapping and line cuts used to identify electrically tunable magnetic states [18]. Copyright 2018, Springer Nature. (b) CrSBr/hexagonal boron nitride (hBN)/graphene (Gr)/$CuInP_2S_6$ (CIPS) floating-gate heterostructure and scanning electron microscopy image of the device used to examine nonvolatile electrical control [138]. Copyright 2024, American Chemical Society. (c) Band diagrams showing charge trapping and detrapping in the graphene floating gate under opposite gate-voltage sweeps [138]. Copyright 2024, American Chemical Society. (d) Rhombohedral and monoclinic stacking structures of bilayer $CrI_3$, providing the crystallographic basis for pressure-tunable interlayer magnetism [20]. Copyright 2019, Springer Nature. (e) Pressure-cell device schematic used to tune the magnetic state of a $CrI_3$ tunnel junction [20]. Copyright 2019, Springer Nature. (f) Pressure-dependent tunneling-current traces showing magnetic-state switching under applied pressure [20]. Copyright 2019, Springer Nature. (g) Ionically gated $Fe_3GeTe_2$ device data, including device geometry, conductance modulation, phase diagram, and anomalous Hall loops used to analyze gate-enhanced ferromagnetism [14]. Copyright 2018, Springer Nature.

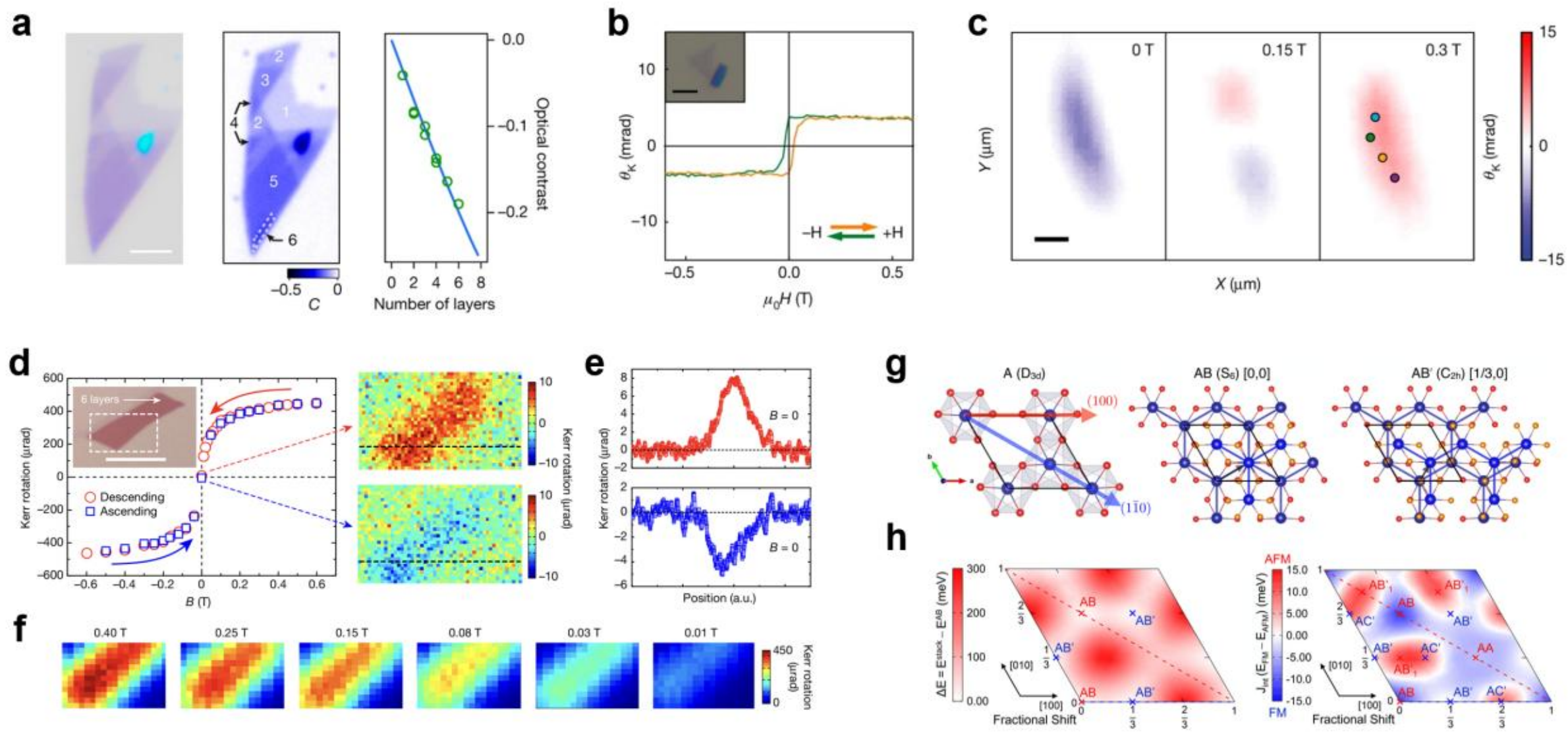

**Figure 6. Optical detection of atomically thin magnetism and stacking-controlled interlayer exchange.** (a) Optical image, contrast map, and layer-number calibration of exfoliated $CrI_3$ flakes [11]. Copyright 2017, Springer Nature. (b) Magneto-optical Kerr effect (MOKE) hysteresis loop of monolayer $CrI_3$, showing out-of-plane ferromagnetic order in the monolayer limit [11]. Copyright 2017, Springer Nature. (c) Field-dependent MOKE maps of $CrI_3$ that spatially resolve magnetic switching [11]. Copyright 2017, Springer Nature. (d) $Cr_2Ge_2Te_6$ Kerr-rotation hysteresis and remanent magnetic response in atomically thin flakes [12]. Copyright 2017, Springer Nature. (e) Kerr-rotation line profiles resolving local magnetic contrast in $Cr_2Ge_2Te_6$ [12]. Copyright 2017, Springer Nature. (f) Field-dependent Kerr images of a single-domain six-layer $Cr_2Ge_2Te_6$ flake [12]. Copyright 2017, Springer Nature. (g) $CrI_3$ stacking configurations, including A, AB, and AB′ registries, considered in models of stacking-dependent magnetism [16]. Copyright 2018, American Chemical Society. (h) Calculated stacking-energy and interlayer-exchange maps for bilayer $CrI_3$, linking lateral registry to ferro- or antiferromagnetic coupling [16]. Copyright 2018, American Chemical Society.

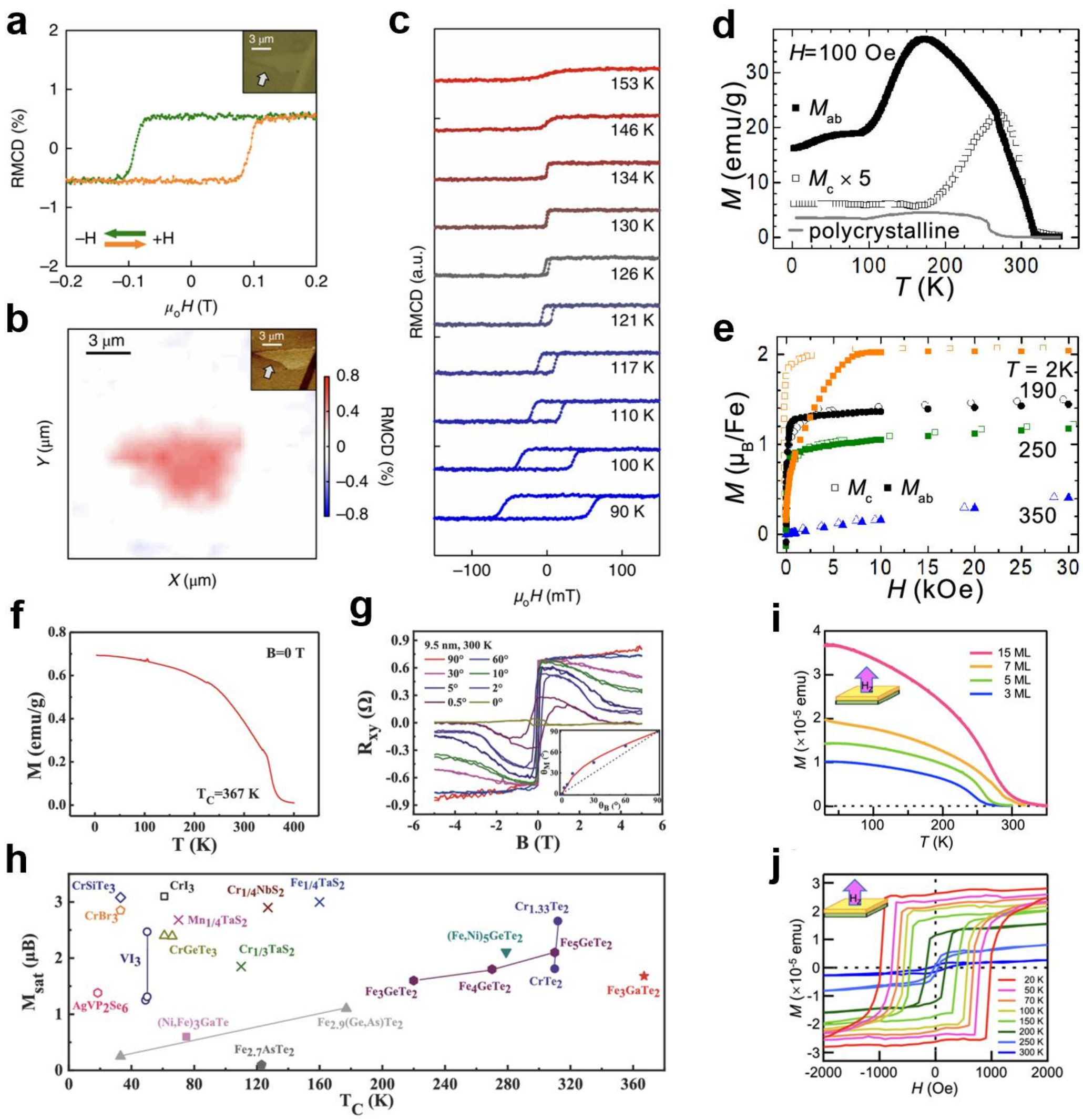

**Figure 7. Representative layered and ultrathin ferromagnets, including high-temperature Fe- and Cr-based systems.** (a) Reflective magnetic circular dichroism (RMCD) hysteresis loop of monolayer $Fe_3GeTe_2$, with the optical image inset identifying the measured flake [13]. Copyright 2018, Springer Nature. (b) RMCD spatial map of monolayer $Fe_3GeTe_2$, showing magnetic contrast over the flake area [13]. Copyright 2018, Springer Nature. (c) Temperature-dependent RMCD loops used to track the persistence of ferromagnetic switching in $Fe_3GeTe_2$ [13]. Copyright 2018, Springer Nature. (d) $Fe_5GeTe_2$ magnetization along in-plane and out-of-plane directions as a function of temperature [88]. Copyright 2019, American Chemical Society. (e) $Fe_5GeTe_2$ magnetic-field sweeps at selected temperatures, showing anisotropic magnetization and high-temperature magnetic response [88]. Copyright 2019, American Chemical Society. (f) Temperature-dependent magnetization of $Fe_3GaTe_2$, showing a high Curie temperature [89]. Copyright 2022, Springer Nature. (g) Angle-dependent anomalous Hall response in $Fe_3GaTe_2$, reflecting strong perpendicular magnetic anisotropy [89]. Copyright 2022, Springer Nature. (h) Comparison of saturation magnetization and Curie temperature across representative vdW magnets, positioning $Fe_3GaTe_2$ among high-temperature candidates [89]. Copyright 2022, Springer Nature. (i) Thickness-dependent magnetization of epitaxial $CrTe_2$ films [90]. Copyright 2021, Springer Nature. (j) Temperature-dependent *M–H* loops of $CrTe_2$ films under perpendicular field, tracking hysteretic magnetic response across a broad temperature range [90]. Copyright 2021, Springer Nature.

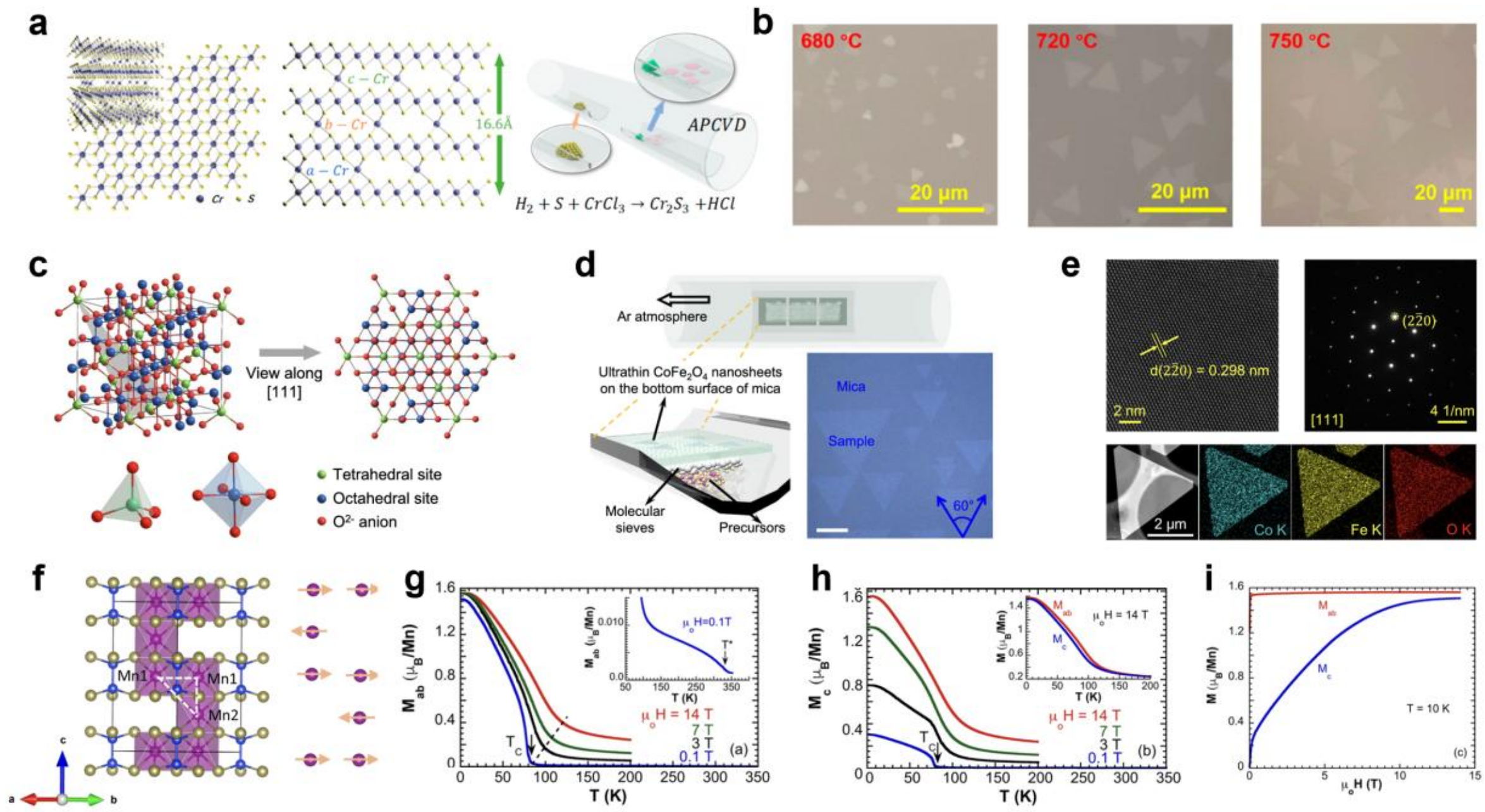

**Figure 8. Layered and ultrathin magnetic systems beyond the prototypical $CrI_3$ and Fe-Ge-Te families.** (a) $Cr_2S_3$ crystal structure and atmospheric-pressure chemical vapor deposition (APCVD) growth schematic for sub-millimeter-scale ultrathin flakes [97]. Copyright 2019, American Chemical Society. (b) Optical micrographs of $Cr_2S_3$ crystals grown at different temperatures, showing temperature-dependent morphology [97]. Copyright 2019, American Chemical Society. (c) $CoFe_2O_4$ spinel crystal structure viewed along the [111] crystallographic direction [70]. Copyright 2022, Springer Nature. (d) Chemical vapor deposition (CVD) setup and optical image of ultrathin $CoFe_2O_4$ nanosheets grown on mica [70]. Copyright 2022, Springer Nature. (e) High-resolution transmission electron microscopy (HRTEM), selected-area electron diffraction (SAED), and transmission electron microscopy energy-dispersive X-ray spectroscopy (TEM-EDS) maps showing crystallinity and elemental distribution in $CoFe_2O_4$ nanosheets [70]. Copyright 2022, Springer Nature. (f) $Mn_3Si_2Te_6$ crystal and magnetic structure, emphasizing the layered ferrimagnetic arrangement [100]. Copyright 2021, American Physical Society. (g) $Mn_3Si_2Te_6$ in-plane magnetization as a function of temperature at selected magnetic fields [100]. Copyright 2021, American Physical Society. (h) Out-of-plane magnetization of $Mn_3Si_2Te_6$ as a function of temperature [100]. Copyright 2021, American Physical Society. (i) Isothermal magnetization of $Mn_3Si_2Te_6$ along in-plane and out-of-plane directions, showing magnetic anisotropy [100]. Copyright 2021, American Physical Society.

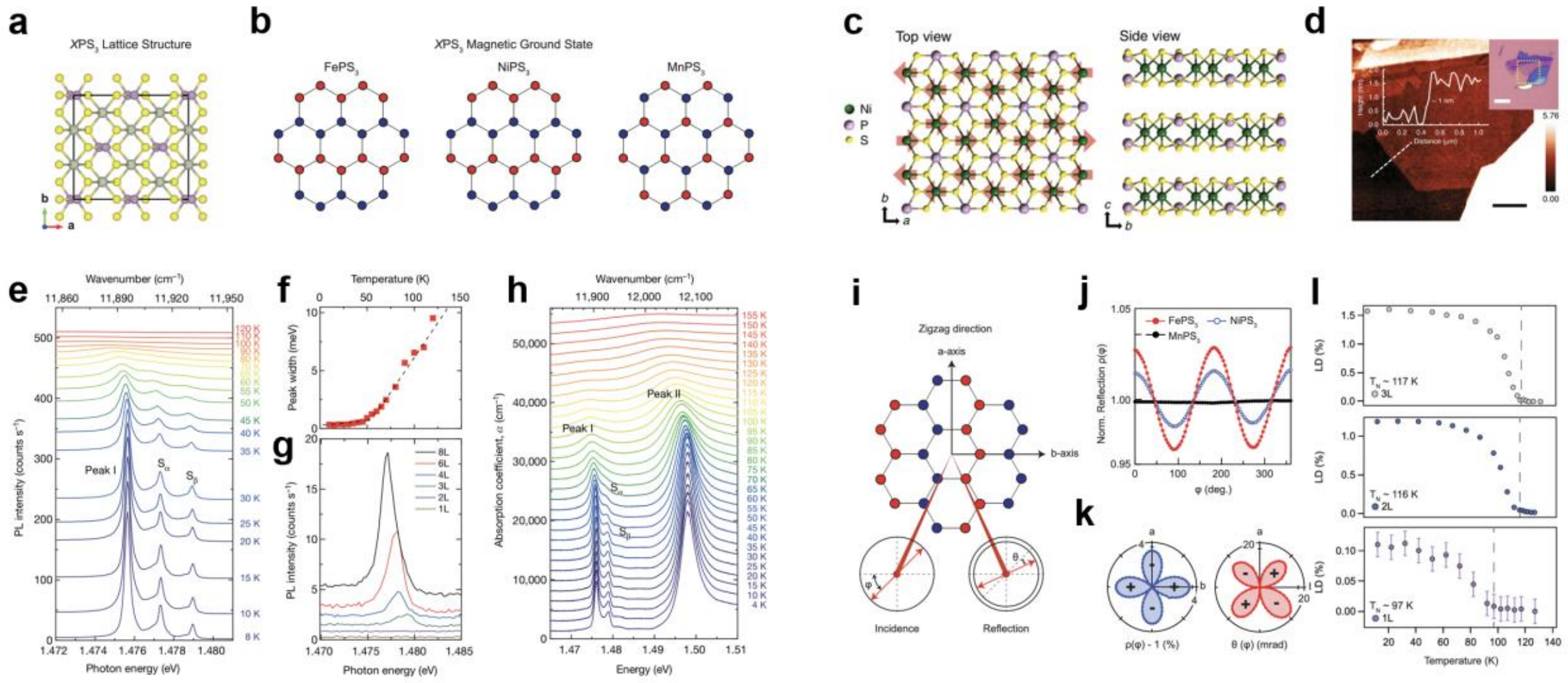

**Figure 9. Antiferromagnetic $MPX_3$ materials and optical probes of magnetic order and symmetry.** (a) Generalized lattice structure for $FePS_3$, $NiPS_3$, and $MnPS_3$, showing the layered honeycomb framework common to these phosphorus trisulfide antiferromagnets [119]. Copyright 2021, American Chemical Society. (b) Magnetic ground states of $FePS_3$, $NiPS_3$, and $MnPS_3$, illustrating the spin patterns available within the $MPX_3$ family [119]. Copyright 2021, American Chemical Society. (c) Top- and side-view $NiPS_3$ crystal and spin structures [118]. Copyright 2019, Springer Nature. (d) Atomic force microscopy image, optical image, and height profile of an exfoliated monolayer $NiPS_3$ flake [118]. Copyright 2019, Springer Nature. (e) Temperature-dependent photoluminescence spectra of $NiPS_3$ [108]. Copyright 2020, Springer Nature. (f) Temperature-dependent excitonic peak width in $NiPS_3$ [108]. (g) Layer-dependent photoluminescence intensity spectra in $NiPS_3$ [108]. Copyright 2020, Springer Nature. (h) Temperature-dependent absorption spectra used to track excitonic features through the magnetic-ordering temperature range [108]. Copyright 2020, Springer Nature. (i) Normal-incidence optical geometry for $FePS_3$ with in-plane polarization referenced to the zigzag direction [119]. Copyright 2021, American Chemical Society. (j) Angle-dependent reflectance anisotropy showing the optical signature of zigzag antiferromagnetic order [119]. Copyright 2021, American Chemical Society. (k) Polar plots of normalized reflectance anisotropy and polarization rotation as functions of light polarization angle [119]. Copyright 2021, American Chemical Society. (l) Temperature-dependent linear dichroism in trilayer, bilayer, and monolayer $FePS_3$, resolving thickness-dependent antiferromagnetic transitions [119]. Copyright 2021, American Chemical Society.

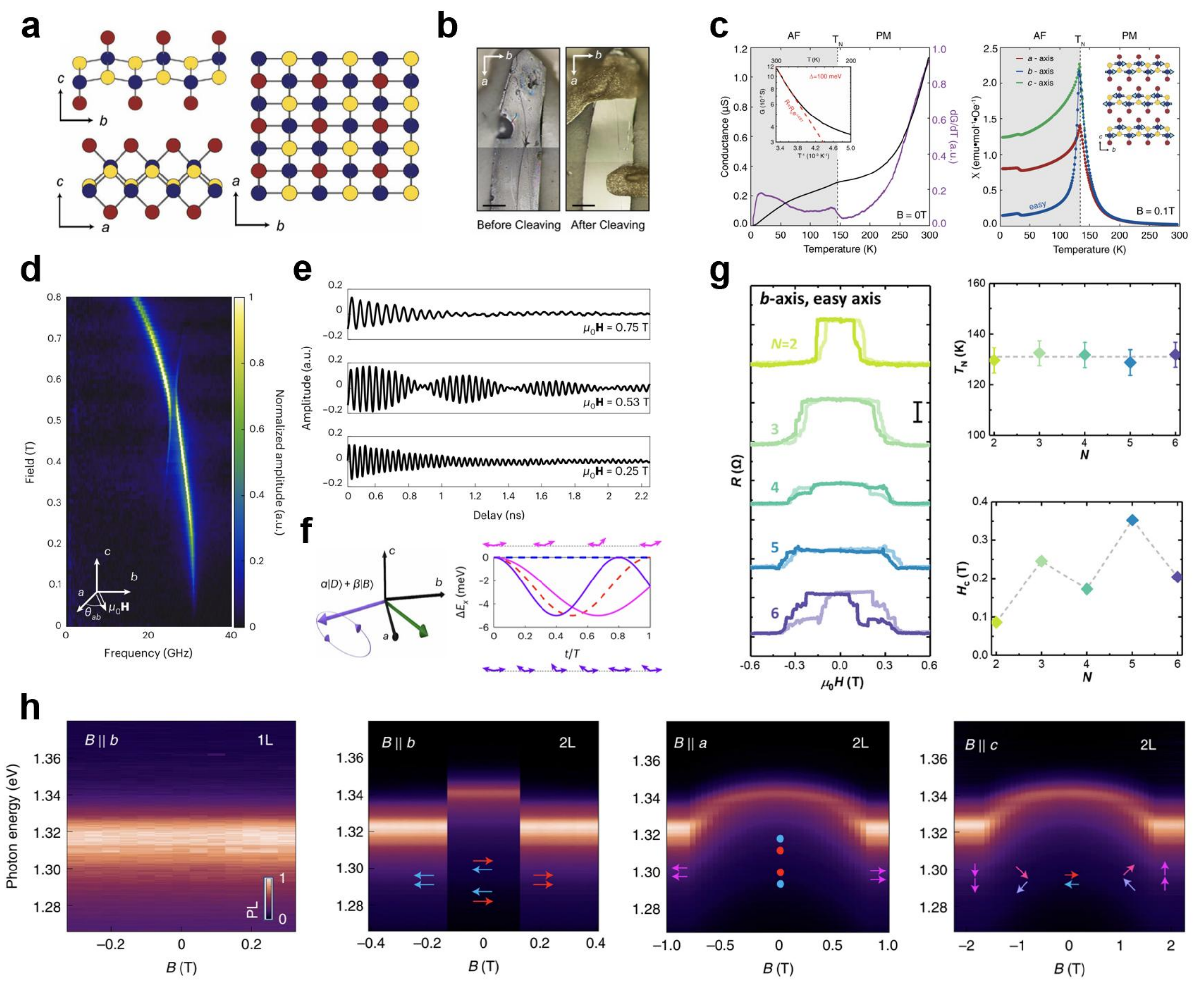

**Figure 10. CrSBr as a layer-dependent air-stable anisotropic magnetic semiconductor.** (a) CrSBr crystal structure viewed along different crystallographic directions, showing the layered orthorhombic lattice and anisotropic bonding [109]. Copyright 2020, John Wiley and Sons. (b) Optical and photographic images of CrSBr before and after cleaving, illustrating mechanical exfoliation from bulk crystals [109]. Copyright 2020, John Wiley and Sons. (c) Temperature-dependent conductance and magnetic susceptibility of CrSBr, showing semiconducting transport and the bulk A-type magnetic transition [109]. Copyright 2020, John Wiley and Sons. (d) Field-frequency map of coherent magnon modes in CrSBr [110]. Copyright 2022, Springer Nature. (e) Time-domain oscillations of coherent magnons under selected magnetic fields [110]. Copyright 2022, Springer Nature. (f) Schematic of a hybridized bright–dark magnon mode and calculated exciton-energy modulation during its precession [110]. Copyright 2022, Springer Nature. (g) Layer-dependent magnetoresistance of CrSBr devices along the magnetic easy axis [122]. Copyright 2022, American Chemical Society. (h) Magnetophotoluminescence spectra of monolayer CrSBr under a field along *b* and bilayer CrSBr under fields along *b*, *a*, and *c* [124]. Copyright 2021, Springer Nature.

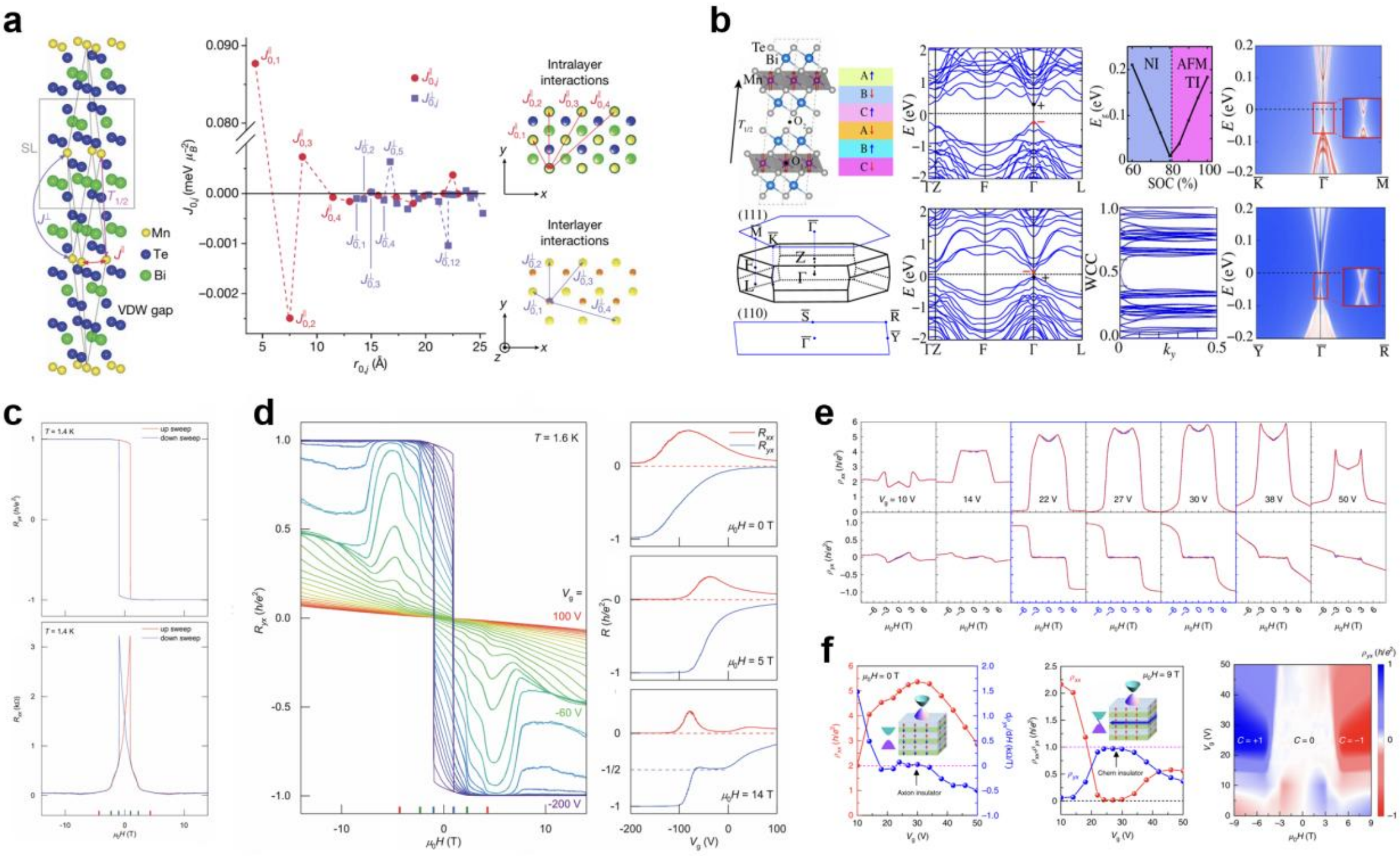

**Figure 11. Magnetic topology and gate-tunable transport in layered $MnBi_2Te_4$.** (a) $MnBi_2Te_4$ septuple-layer crystal structure and calculated exchange interactions, distinguishing intralayer and interlayer magnetic coupling [112]. Copyright 2019, Springer Nature. (b) Band structures, topological phase diagram, and surface-state calculations for antiferromagnetic $MnBi_2Te_4$ [140]. Copyright 2019, The American Association for the Advancement of Science. (c) Quantized Hall resistance and corresponding longitudinal resistance traces from a five-septuple-layer $MnBi_2Te_4$ device in the quantum anomalous Hall regime [113]. Copyright 2020, The American Association for the Advancement of Science (d) Gate-tuned Hall and longitudinal resistance in five-septuple-layer $MnBi_2Te_4$, showing electrical control of the quantum anomalous Hall response [113]. Copyright 2020, The American Association for the Advancement of Science. (e) Gate-dependent transport of six-septuple-layer $MnBi_2Te_4$ across magnetic-field and gate-voltage windows [144]. Copyright 2020, Springer Nature. (f) Gate-dependent transport curves and a Hall-resistivity phase map showing axion-insulator-like and Chern-insulator regimes [144]. Copyright 2020, Springer Nature.

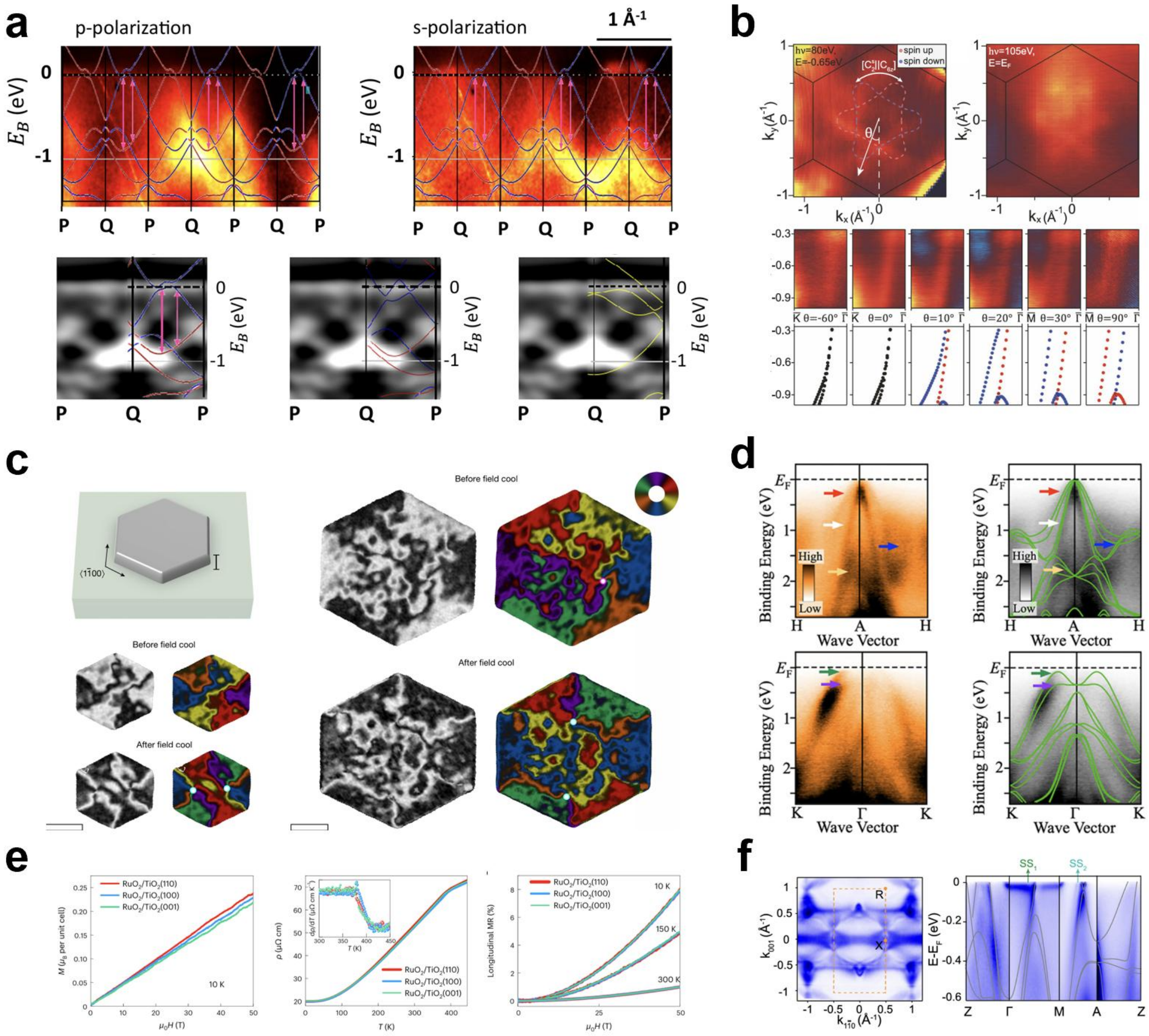

**Figure 12. Experimental benchmarks and contested diagnostic evidence for altermagnetism.** (a) Soft-X-ray angle-resolved photoemission spectroscopy (SX-ARPES) of a 30 nm epitaxial CrSb film [168]. Copyright 2024, Springer Nature. (b) In-plane spin-integrated ARPES maps and calculated spin-resolved bands of bulk CrSb single crystals [169]. Copyright 2025, Springer Nature. (c) Domain imaging and control in a 30 nm epitaxial MnTe film [171]. Copyright 2024, Springer Nature. (d) ARPES evidence of band splitting in bulk MnTe single crystals [173]. Copyright 2024, American Physical Society. (e) $RuO_2$ magnetization and magnetotransport measurements that provide supporting but sample-dependent diagnostic evidence [163]. Copyright 2022, Springer Nature. (f) $RuO_2$ electronic-structure measurements that form part of the conflicting experimental evidence [165]. Copyright 2024, American Physical Society.

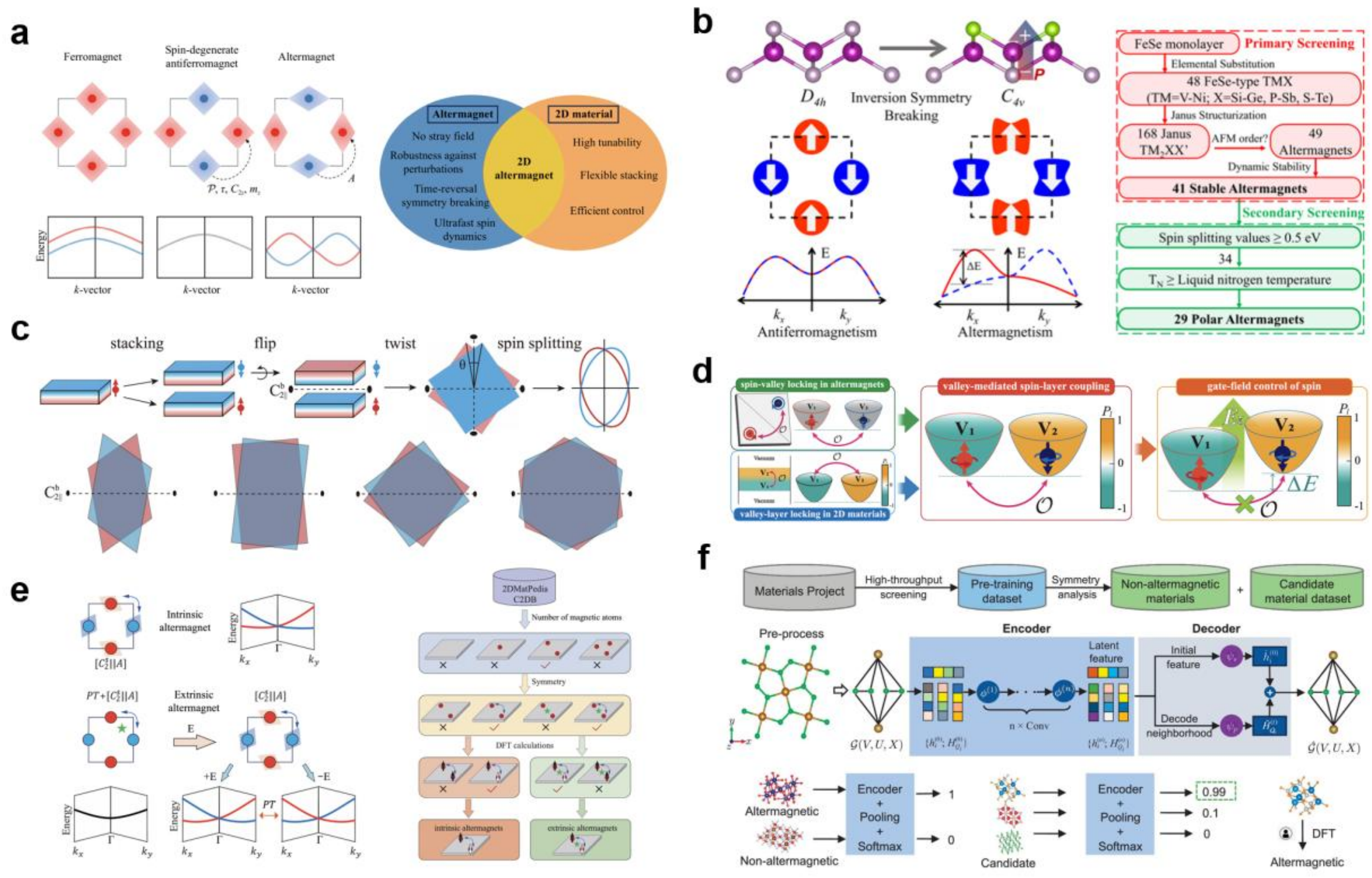

**Figure 13. Design routes for candidate two-dimensional altermagnetic platforms**. (a) Representative collinear ferromagnetic, spin-degenerate antiferromagnetic, and altermagnetic states, with schematic nonrelativistic band structures and proposed advantages of 2D altermagnets [187]. Copyright 2026, Higher Education Press. (b) Polarity-driven altermagnetism in Janus FeSe-type monolayers and the corresponding screening workflow [206]. Copyright 2025, American Chemical Society. (c) Stacking, flipping, and twisting routes for generating altermagnetic spin splitting in magnetic van der Waals bilayers [201]. Copyright 2024, American Physical Society. (d) Spin-valley and valley-layer locking in altermagnets, including gate-field control of spin polarization [203]. Copyright 2024, American Physical Society. (e) Intrinsic and extrinsic 2D altermagnets, electric-field-induced switching of spin splitting, and a conventional high-throughput screening workflow for 2D altermagnetic candidates [209]. Copyright 2024, American Chemical Society. (f) AI-accelerated discovery workflow for altermagnetic candidates [218]. Copyright 2025, Oxford University Press.

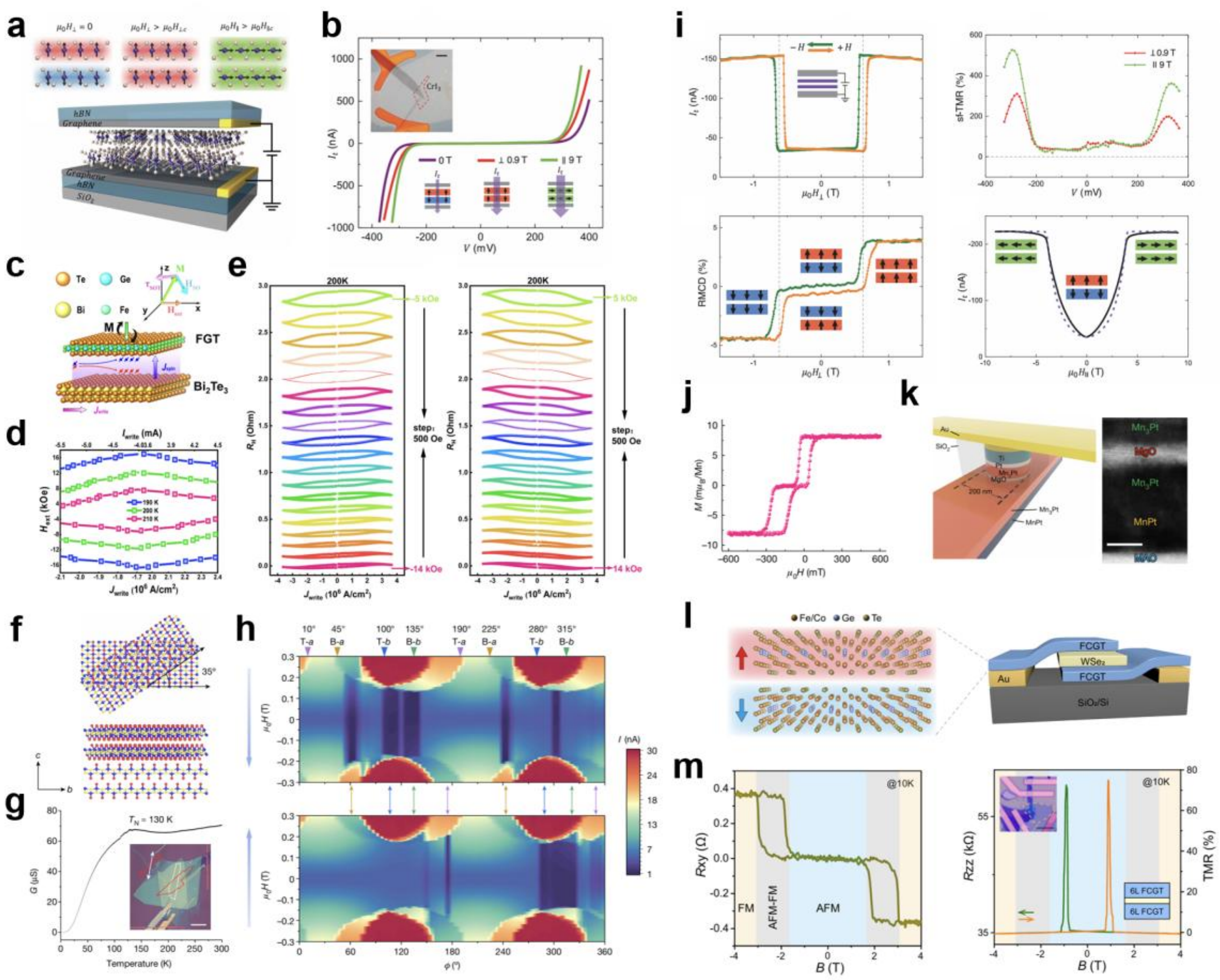

**Figure 14. Spin-filter, spin–orbit torque, and tunneling devices based on two-dimensional magnets and benchmark systems with compensated magnetic order.** (a) $CrI_3$ spin-filter states and graphene/hexagonal boron nitride (hBN)/$CrI_3$ tunnel-junction geometry [23]. Copyright 2018, The American Association for the Advancement of Science. (b) Bilayer $CrI_3$ current–voltage characteristics at selected magnetic fields [23]. Copyright 2018, The American Association for the Advancement of Science. (c) $Bi_2Te_3$/$Fe_3GeTe_2$ spin–orbit torque device and switching geometry [229]. Copyright 2023, Springer Nature. (d) External-field versus writing-current-density switching diagram at selected temperatures [229]. Copyright 2023, Springer Nature. (e) Current-induced switching at 200 K [229]. Copyright 2023, Springer Nature. (f) A 35°-twisted CrSBr bilayer/bilayer junction [242]. Copyright 2024, Springer Nature. (g) Temperature-dependent device conductance [242]. Copyright 2024, Springer Nature. (h) Field-orientation-dependent tunneling-current maps [242]. Copyright 2024, Springer Nature. (i) Double spin-filter junction based on bilayer $CrI_3$, showing tunneling current, reflective magnetic circular dichroism (RMCD), tunneling magnetoresistance (TMR), and field dependence [23]. Copyright 2018, The American Association for the Advancement of Science. (j) Out-of-plane magnetization of the $Mn_3Pt$ layers in the multilayer stack [231]. Copyright 2023, Springer Nature. (k) Schematic and cross-sectional transmission electron microscopy (TEM) image of the $Mn_3Pt$/MgO/$Mn_3Pt$/MnPt junction [231]. Copyright 2023, Springer Nature. (l) A-type antiferromagnetic $(Fe_{0.6}Co_{0.4})_5GeTe_2$ (FCGT) and FCGT/$WSe_2$/FCGT junction [232]. Copyright 2025, Springer Nature. (m) Hall and tunnel transport, including $R_{xy}$, $R_{zz}$, and TMR [232]. Copyright 2025, Springer Nature.